\documentclass[prd,showpacs,preprintnumbers,onecolumn,superscriptaddress,floatfix]{revtex4-2}

\usepackage[utf8]{inputenc}
\usepackage{bm,amsmath,amssymb}
\usepackage{graphicx}
\usepackage[normalem]{ulem}
\usepackage{comment}
\usepackage{color}
\usepackage[colorlinks=true,
  citecolor=teal,
  linkcolor=,
  urlcolor=teal]{hyperref}
\usepackage{multirow}
\usepackage{caption}
\usepackage{subcaption}
\usepackage{lineno}
\usepackage{soul}
\usepackage{float}
\usepackage{tikz}
\usepackage{MnSymbol}
\usepackage{bigints}
\usetikzlibrary{shapes.geometric, arrows.meta, positioning}
\usepackage{xcolor}
\usepackage[nameinlink]{cleveref}

\crefformat{equation}{\textcolor{teal}{Eq.}(#2#1#3)}
\crefformat{figure}{\textcolor{teal}{Fig.}\,#2#1#3}

\tikzstyle{startstop} = [rectangle, rounded corners, minimum width=3cm, minimum height=1cm, text centered, draw=black, fill=red!30]
\tikzstyle{process} = [rectangle, minimum width=3cm, minimum height=1cm, text centered, draw=black, fill=orange!30]
\tikzstyle{decision} = [rectangle, minimum width=3cm, minimum height=1cm, text centered, draw=black, fill=green!30, aspect=2]
\tikzstyle{arrow} = [thick, ->, >=Stealth]
\tikzstyle{arrow} = [thick, ->, >=Stealth]
\let\oldalign\align
\def\align{\linenomath\oldalign}

\begin{document}

\title{Lattice study of spin interactions between heavy quarks in the quark-gluon plasma}

\author{Dibyendu Bala}
\email{dibyendu.bala@physik.uni-bielefeld.de}
\affiliation{Fakult\"{a}t f\"{u}r Physik, Universit\"{a}t Bielefeld, D-33615 Bielefeld, Germany}
\author{Olaf Kaczmarek}
\email{okacz@physik.uni-bielefeld.de}
\affiliation{Fakult\"{a}t f\"{u}r Physik, Universit\"{a}t Bielefeld, D-33615 Bielefeld, Germany}
\author{Sayantan Sharma}
\email{sayantans@imsc.res.in}
\affiliation{The Institute of Mathematical Sciences, Chennai 600113, India}
\affiliation{Homi Bhabha National Institute, Training School Complex, Anushaktinagar, Mumbai 400094, India}
\author{Swagatam Tah}
\email{swagatamt@imsc.res.in}
\affiliation{The Institute of Mathematical Sciences, Chennai 600113, India}
\affiliation{Homi Bhabha National Institute, Training School Complex, Anushaktinagar, Mumbai 400094, India}

\begin{abstract}

We calculate the spin-dependent potential, which is the $\mathcal{O}(1/M^2)$ correction term to the thermal potential between a static quark-antiquark pair within non-relativistic QCD. At leading 
order in hard thermal loop perturbation theory, we show that this spin-dependent potential has an imaginary part which is different in 
magnitude for pseudoscalar and vector quarkonium states.  For the first time, we extract the imaginary part non-perturbatively using 
lattice techniques, in the deconfined phase of quenched QCD at $T\sim 470$ MeV, after performing a continuum estimation and subsequent 
renormalization. We have found that the spin-dependent potential in the quark-gluon plasma phase is complex, and its imaginary part has a
remarkably significant contribution over the thermal static potential for charmonium states. Consequences of this thermal spin-dependent potential on the quarkonium spectral functions are also discussed.
\end{abstract}

\maketitle


\section{Introduction}

The deconfined phase of nuclear matter, the quark-gluon plasma (QGP) provides a unique system to study 
both the perturbative as well as non-perturbative aspects of QCD~\cite{Gross:1980br,Shuryak:1993kg,Wang:2025lct}. 
Such a phase of matter is believed to exist in the early universe and an extensive research went into designing
favorable conditions for its re-creation in heavy-ion collision experiments at RHIC and 
LHC~\cite{Bjorken:1982qr}. The formation of QGP in experiments is only known indirectly~\cite{McLerran:1986zb} 
through observations like jet quenching~\cite{Bjorken1982}, collective flow of hadrons~\cite{Ollitrault:1992bk}, 
photon and di-lepton production rates~\cite{Shuryak:1978ij} and strangeness enhancement~\cite{Koch:1986ud}. Among these, the
dynamics of quarkonia, the bound states of a heavy quark and antiquark, serves as an excellent probe 
of the formation time and the dynamics of QGP formed in a heavy-ion collision event~\cite{Matsui1986}. This is 
possible since quarkonia are produced at very early times, $\lesssim 0.1$ fm, due to perturbative hard processes 
which allow them to propagate during the entire epoch of the evolution of QGP~\cite{Mocsy:2013syh}. 
As a result, their in-medium interactions modify their survival probability~\cite{Matsui1986} and 
yields. Quarkonium yields observed in heavy-ion collisions scaled by the number of binary 
collisions thus differ from the expectation based on proton-proton 
collisions~\cite{CMS:2016rpc,STAR:2022rpk,ALICE:2022jeh}.

From a theoretical point of view, understanding these modifications requires 
access to the real-time dynamics of quarkonia in a thermal medium. Since the heavy-quark 
mass $M$ is larger than both the strong-interaction scale $\Lambda_{\mathrm{QCD}}$ and the 
temperature of the QGP, which decides its typical momentum, one can perform a systematic expansion 
of the heavy-quark Hamiltonian as a series in $1/M$. Within this so-called NRQCD framework~\cite{Thacker:1990bm}, 
the quarkonium correlation functions can be calculated from a non-relativistic Schr\"odinger equation 
in the presence of an interquark potential~\cite{Brambilla2005,Brambilla2000}. 
At zero temperature, this potential can be determined non-perturbatively from large Wilson loops, 
consisting of a string-like term dominating at large separations and a Coulomb-like term at short 
distances~\cite{Bali:1992ab,Bali:2000gf,Booth:1992bm,Glassner:1996xi,PhysRevD.71.114510}. 
The LO potential can describe the gross features of the charmonium energy levels and their degeneracies~\cite{Voloshin:2007dx} at zero temperature.

The potential was first calculated at LO in strong coupling expansion in resummed
hard thermal loop (HTL) perturbation theory, where it was shown that, unlike at zero temperature, 
it develops an imaginary part at finite temperatures~\cite{Laine:2006ns}. The real part of the 
potential is responsible for color screening in the QGP, whereas its imaginary part is responsible 
for Landau damping, due to scattering of the heavy quark-antiquark pair with in-medium hard partons. 
Later, within the pNRQCD formalism at weak coupling, it was shown that the imaginary part also receives 
contributions due to gluodissociation, where a quarkonium state absorbs a thermal gluon, thereby transitioning 
to a color-octet state~\cite{Brambilla:2008cx}. 
This imaginary part of the thermal potential at finite temperature makes its extraction, using lattice 
QCD  considerably more involved. On the lattice, correlation function of Wilson lines are computed in 
Euclidean time, whereas the origin of an imaginary potential at finite temperatures in continuum QCD is due 
to real-time dynamics. Consequently, one needs to perform an analytic continuation of the correlation 
function calculated on a discrete set of points along the Euclidean time direction to continuous Minkowski 
time, which is a well-known ill-posed problem~\cite{Rothkopf:2011db}. Additional physics-motivated inputs 
are therefore required to ensure the uniqueness of analytic continuation. An extensive body of work has been 
performed along these lines, using Bayesian methods and physics-driven analytic continuation strategies to 
extract the thermal potential~\cite{Rothkopf:2011db, Burnier:2014ssa,Burnier:2015tda, Bala:2019cqu, Ali:2025iux}. These studies consistently 
indicate sizable non-perturbative contributions to the thermal potential in QGP. The in-medium behavior of 
quarkonium states is encoded in their spectral functions, which can be obtained by solving the 
Schrödinger equation with the corresponding thermal potential~\cite{Burnier:2007qm}. The correlators 
reconstructed from these spectral functions compare well with the correlators calculated directly on the lattice~\cite{Burnier:2017bod, Ali:2025iux}.

The energies of the pseudoscalar and vector quarkonium states obtained by solving the Schrödinger equation with 
the LO thermal potential are degenerate. In order to describe the physically realistic scenario wherein these 
states are non-degenerate, one would need to go beyond the LO, i.e static limit and estimate the  
$1/M^\alpha, \alpha \geq 1$ terms in the potential. The term that breaks this degeneracy appears 
at $\mathcal{O}(1/M^2)$, which denotes spin-dependent interaction. At zero temperature, this spin-dependent interaction 
term has been computed non-perturbatively in quenched QCD with different colors~\cite{Bali:1996cj, Bali:1997am, Koma:2006fw} and has been used to explain the hyperfine splitting observed between quarkonium 
states~\cite{Voloshin:2007dx, Laschka:2011zr}. In contrast, the temperature dependence of this spin-dependent potential has not yet been investigated from first principles. We will, for the first time, 
extract this spin-dependent potential in thermal QCD using lattice techniques. This would enable us 
to quantify thermal effects due to the QGP on the spectral functions corresponding to quarkonium states in 
different quantum number channels. Accurately quantifying the thermal mass shifts and decay widths in the pseudoscalar and vector quarkonium channels is of phenomenological importance, especially for charmonium, where the effects of spin-dependent corrections are expected to be larger. The spin-dependent 
interaction will also play an important role in understanding quarkonium polarization observed in heavy-ion collisions~\cite{ALICE:2022dyy}. 

The paper is organized as follows: In the next section, we outline the essential steps in the derivation of the 
spin-dependent potential from Wilson loop correlators with chromomagnetic field insertions, starting 
from the NRQCD Lagrangian. In \cref{sec:SpinPert}, we calculate this potential at zero temperature using the transfer 
matrix approach and at finite temperatures using HTL perturbation theory and also within the pNRQCD effective field 
theory. In \cref{sec:LatticeDetails}, we discuss the details of our lattice simulations, followed by a detailed 
description of our procedure to extract the renormalized non-perturbative spin-dependent potential in the 
subsequent section. Next, we present our final results on the real and imaginary parts of the renormalized 
spin-dependent potential and discuss their implications for quarkonium states in QGP by calculating their 
spectral functions. Finally, we summarize our work and outline some future directions that can be further explored.

\section{spin-dependent potential}
\label{sec:Theo}
In this section, we begin by outlining the derivation of the spin-dependent potential between 
a static quark-antiquark pair separated by distance $r$ within the NRQCD formalism at temperature 
$T$. Since the mass of the heavy quark is much larger than the relevant scales, $ M \gg T,\Lambda_{\text{QCD}},$ 
its contribution to the QCD Lagrangian in Euclidean time can be written as
\begin{equation}
L_h=\theta^{\dagger}\left(D_\tau+M - c_2(\mu)\frac{ \vec  D^2}{2 M} + c_{B}(\mu,M) \frac{g(\mu)}{2 M} \vec{\sigma} 
\cdot \vec{B}\right)\theta+\chi^{\dagger}\left(D_\tau-M+c_2(\mu)\frac{ \vec  D^2}{2 M} - c_{B}(\mu,M) \frac{g(\mu)}{2 M} \vec{\sigma} \cdot \vec{B}\right)\chi+...,
\label{hql}
\end{equation}
where $\theta$ and $\chi$ are two components of the heavy quark field. $\vec{B}, \vec{E}$ are the 
chromomagnetic and chromoelectric fields respectively and $D_\tau, \vec{D}$ represents the covariant 
derivatives along the temporal and spatial directions. The coefficients $c_{2}(\mu)$ and 
$c_{B}(\mu,M)$ are the Wilson coefficients which are obtained by matching any physical observable calculated with effective theory and with QCD at a scale $\mu\sim M$. Reparameterization 
invariance of the effective theory allows us to set $c_2(\mu)=1$~\cite{Luke:1992cs}.  
We will now derive the spin correlation functions in four-dimensional Euclidean space-time which can 
be directly implemented in lattice calculations to extract the spin-dependent potential. A heavy quark-antiquark pair ($q\bar{q}$) in the quantum number 
channel denoted by $\Gamma$, separated by a distance $r=|\vec{x}-\vec{y}|$ and at time $\tau$ can be 
created from the vacuum by the operator 
$\hat{\mathcal{P}}^\dagger_\Gamma(r,\tau) =\chi^{\dagger}(\vec{x},\tau)\Gamma\, U(\vec{x}-\vec{y},\tau)\,\theta(\vec{y},\tau)$.
Here $U(\vec x-\vec y,\tau)$ is the gauge link that connects the quark-antiquark pair and makes the operator 
gauge-invariant. The correlation function of these pair states is defined as the thermal average over 
the vacuum expectation values  
\begin{equation}
C_\Gamma(r,\tau)=\langle P^{\dagger}_\Gamma(r,\tau) P_\Gamma(r,0)\rangle_T
\end{equation}
This correlation function can be calculated starting from the Lagrangian in \cref{hql} and using the 
following propagators,
\begin{equation}
\begin{aligned}
\langle \theta_\alpha(\mathbf{x},\tau)\, \theta_\beta^\dagger(\mathbf{y},0) \rangle
&= \delta^{(3)}(\mathbf{x}-\mathbf{y})\, U_{\alpha\beta}(\tau,0)\, e^{-\tau M} \, , \\
\langle \theta_\alpha(\mathbf{x},0)\, \theta_\beta^\dagger(\mathbf{y},\tau) \rangle
&= -\,\delta^{(3)}(\mathbf{x}-\mathbf{y})\, U_{\alpha\beta}(\beta,\tau)\, e^{-(\beta-\tau)M} \, , \\
\langle \chi_\alpha(\mathbf{x},\tau)\, \chi_\beta^\dagger(\mathbf{y},0) \rangle
&= \delta^{(3)}(\mathbf{x}-\mathbf{y})\, U_{\alpha\beta}^\dagger(\beta,\tau)\, e^{-(\beta-\tau)M} \, , \\
\langle \chi_\alpha(\mathbf{x},0)\, \chi_\beta^\dagger(\mathbf{y},\tau) \rangle
&= -\,\delta^{(3)}(\mathbf{x}-\mathbf{y})\, U_{\alpha\beta}^\dagger(\tau,0)\, e^{-\tau M} \, .
\end{aligned}
\label{propagators}
\end{equation}
In the static limit the leading contribution to the correlator is
\begin{equation}
C_\Gamma(r,\tau)=\exp(-2\,M\,\tau)\,\langle \text{Tr}_c[ W(r,\tau)]\rangle_T\equiv\exp(-2\,M\,\tau)\,W_T(r,\tau)~,
\label{eqn:correlator}
\end{equation}
where $W_T(r,\tau)$ is the thermal average of the trace of the Wilson loop operator and the trace is over 
color indices. However, in the static limit there is no distinction between different quantum channels. The spin-dependent term which appears at order $1/M^2$ breaks this degeneracy. This spin-dependent potential can be 
calculated by considering the following spin-dependent chromomagnetic part of the NRQCD Lagrangian, 
\begin{equation}
-c_{B}(\mu,M)\frac{g(\mu)}{2M}\left[\theta^{\dagger}\left(\vec \sigma \cdot B\right)\theta-\chi^{\dagger}\left( \vec \sigma \cdot B\right)\chi\right] \nonumber
\end{equation}
and treating this as a small $1/M$ perturbation to the Lagrangian due to static quarks. Using the 
propagators defined in \cref{propagators} one can now derive the correlator in \cref{eqn:correlator}
at $\mathcal{O}(1/M^2)$, 
\begin{align}
C_{\Gamma}(r,\tau)= \exp(-2\,M\,\tau) &\mathrm{Tr}_c[\Gamma^2] W_T(r,\tau)\left[1 +\frac{c^2_{B}(\mu,M)}{4\,M^2}W_\text{BB}(r,\tau,\mu)\right]. 
\label{eqn:correlatorBB} \\
\text{where,} ~~~ W_\text{BB}(r,\tau,\mu) &\equiv \mathcal{X}_{ij}\,W^{\text{int}}_{B_i B_j}(r,\tau,\mu)+ \mathcal{Y}_{ij}\,W^{\text{self}}_{B_iB_j}(r,\tau,\mu) \nonumber
\end{align}
In the above expression for $W_{BB}$, 
$W^{\text{self}}_{B_iB_j}$ corresponds to the self-energy contribution
of the pair due to the $B$ fields acting only on either the quark or the
antiquark, whereas $W^{\text{int}}_{B_iB_j}$ represents the interaction
between the pair in the presence of the chromomagnetic fields. The
indices $i,j$ are contracted with $\mathcal{X}_{ij}$ and
$\mathcal{Y}_{ij}$, which are quantum-channel dependent and are given by
\begin{align}
\mathcal{X}_{ij}
&= \frac{\mathrm{Tr}_d\!\left(\Gamma \sigma_i \Gamma \sigma_j\right)}
{\mathrm{Tr}_d\!\left(\Gamma^2\right)},
\qquad
\mathcal{Y}_{ij}
= \frac{\mathrm{Tr}_d\!\left(\sigma_i \sigma_j \Gamma \Gamma\right)}
{\mathrm{Tr}_d\!\left(\Gamma^2\right)} .
\end{align}
For the $q\bar{q}$ pair in the pseudoscalar channel, $\Gamma = \mathbf{1}$, while for the 
vector state, $\Gamma = \sigma_k, k=1,2,3$. This leads to $\mathcal{X}_{ij}=\delta_{ij}$ and 
$\mathcal{X}_{ij}=-\delta_{ij}/3$ for the pseudo-scalar and vector channels respectively.
In both cases, however, $\mathcal{Y}_{ij}=\delta_{ij}$. As a result, the 
magnetic-field indices in the correlators are contracted by $\delta_{ij}$ and therefore we 
need to consider the following contributions only,  
\begin{figure}[H]
\centering
\begin{minipage}{0.82\textwidth}
\normalsize
\begin{align}
W_{\mathrm{BB}}^{\mathrm{int}}(r,\tau,\mu)
&= 
\int_0^{\tau}\!\!\int_0^{\tau} d\tau_1\, d\tau_2
\frac{
\Big\langle
\mathrm{Tr}_c\!\Big[
\mathcal{T}\, W(r,\tau)\,
g(\mu) B_i(\vec{y},\tau_1)\,
g(\mu) B_i(\vec{x},\tau_2)
\Big]
\Big\rangle_T
}{
W_T(r,\tau)
}\,,
\label{eqn:IntSpin}
\end{align}
\end{minipage}
\hfill
\begin{minipage}{0.12\textwidth}
\normalsize
\centering
\begin{tikzpicture}[scale=1.3, thick]
  \node at (1.3,1.0) {$\tau_1$};
  \node at (1.8,2.0) {$\tau_2$};
  \draw[->, gray, thick] (1,0.5) -- (1,1.5);
  \draw[gray, thick] (1,1.5) -- (1,2.5);
  \draw[->, gray, thick] (2,2.5) -- (2,1.5);
  \draw[gray, thick] (2,1.5) -- (2,0.5);
  \draw[gray, thick] (1,0.5) -- (2,0.5);
  \draw[gray, thick] (1,2.5) -- (2,2.5);
  \fill[orange] (1,1.0) circle (2pt);
  \fill[orange] (2,2.0) circle (2pt);
\end{tikzpicture}
\end{minipage}
\end{figure}
\vspace{-0.5cm}

\begin{figure}[H]
\centering
\begin{minipage}{0.82\textwidth}
\normalsize
\begin{align}
W_{\mathrm{BB}}^{\mathrm{self}}(r,\tau,\mu)
= 
\int_0^{\tau}\!\!\int_0^{\tau} d\tau_1\, d\tau_2
\frac{
\Big\langle
\mathrm{Tr}_c\!\Big[
\mathcal{T}\, W(r,\tau)\,
g(\mu) B_i(\vec{x},\tau_1)\,
g(\mu) B_i(\vec{x},\tau_2)
\Big]
\Big\rangle_T
}{
W_T(r,\tau)
}. \,
\label{eqn:SelfSpin}
\end{align}
\end{minipage}
\hfill
\begin{minipage}{0.12\textwidth}
\centering
\normalsize
\begin{tikzpicture}[scale=1.3, thick]
  \node at (0.87,0.5) {$\chi^\dagger$};
  \node at (2.1,0.5) {$\theta$};
  \node at (1.5,0.2) {$r$};
  \draw[<-, black, thick] (1,0.2) -- (1.4,0.2);
  \draw[->, black, thick] (1.6,0.2) -- (2.0,0.2);
  \node at (2.3,1.5) {$\tau$};
  \draw[<-, black, thick] (2.3,0.5) -- (2.3,1.4);
  \draw[->, black, thick] (2.3,1.6) -- (2.3,2.5);
  \node at (1.3,1.0) {$\tau_1$};
  \node at (1.3,2.0) {$\tau_2$};
  \draw[->, gray, thick] (1,0.5) -- (1,1.5);
  \draw[gray, thick] (1,1.5) -- (1,2.5);
  \draw[->, gray, thick] (2,2.5) -- (2,1.5);
  \draw[gray, thick] (2,1.5) -- (2,0.5);
  \draw[gray, thick] (1,0.5) -- (2,0.5);
  \draw[gray, thick] (1,2.5) -- (2,2.5);
  \fill[orange] (1,1.0) circle (2pt);
  \fill[orange] (1,2.0) circle (2pt);
\end{tikzpicture}
\end{minipage}
\end{figure}

On the right-hand side of the above expressions, the diagrammatic representations
of the spin-dependent correlators are also shown, where the dots represent
chromomagnetic field insertions and the lines represent Wilson lines. In the 
subsequent sections we will, for simplicity, use the notation 
$\iint_0^\tau  \equiv \int_0^{\tau} \int_0^{\tau} d\tau_1\, d\tau_2$.

The potential between a heavy quark and antiquark pair in a given quantum-number
channel is defined only in real time $t$. It can be obtained by analytically
continuing the Euclidean correlator  in~\cref{eqn:correlatorBB} and then
taking the long-time limit,
\begin{equation}
V_{\Gamma}(r)
=
\lim_{t\to\infty}
i\partial_t \log C_{\Gamma}(r,\tau\to it)
=
2\,M+V_{\text{static}}(r)
+\underbrace{\frac{\Phi^{\text{self}}(r)}{4\,M^2}}_{V^{\text{self}}(r)}
-\frac{4}{3}\,\vec{s}_1\cdot\vec{s}_2\,
\underbrace{\frac{\Phi^{\text{int}}(r)}{4\,M^2}}_{V^{\text{int}}(r)}~.
\label{defpot}
\end{equation}
Here $V_{\text{static}}(r)$ is the static potential defined in terms of the
Wilson loop without any chromomagnetic field insertions. The last two terms 
in the right hand side of~\cref{defpot} represents the spin-dependent contribution 
to the potential consisting of the terms $\Phi^{\text{self}}(r)$ and $\Phi^{\text{int}}(r)$, 
which are defined as
\begin{align}
\Phi^{\text{self/int}}(r)
&=
\lim_{t \to \infty}
i\partial_t\,
W^{\text{self/int}}_{\text{BB}}(r,\tau \to it,\mu)\,
c^2_{B}(\mu,M) .
\label{defss}
\end{align}
The factor $\vec{s}_1\cdot\vec{s}_2=-3/4~(1/4)$ for pseudoscalar (vector) 
channels respectively. The functions $\Phi^{\text{self/int}}(r)$ have mass 
dimension three, nevertheless contain all the non-trivial $r$-dependence 
of spin-dependent interactions. We will henceforth refer to these as 
the spin potential, however, to obtain the physical potential which has 
a mass dimension of unity, one needs to normalize by the factor $1/4M^2$.

The goal of this work is to calculate this spin-dependent potential using
non-perturbative lattice techniques. Before doing so, it is useful to understand
some general features of the potential defined in~\cref{defss}, both at zero and
finite temperatures. This will be discussed in the following section.

\section{Properties of the spin-dependent correlation function}
\label{analytic}
In order to perform an analytic continuation described by \cref{defss}, 
it is instructive to understand the analytic properties of the correlator 
itself. In this section we discuss two methods that can be used to understand 
these properties. The first method is based on the transfer matrix formalism and gives us 
an analytic understanding of the properties of the potential arising due to spin 
interactions at zero temperature. At finite temperatures, we use the resummed HTL propagator 
in order to derive the expression of the spin-dependent potential. We then 
also calculate its dependence for the small spatial separation $rT \ll 1$ within an 
effective theory, pNRQCD, which describes the interactions among $q\bar{q}$ states.

\subsection{Zero temperature}

The expression for the interaction part of the spin-dependent correlator in \cref{eqn:IntSpin} can be written 
in the Heisenberg formalism as
\begin{equation}
W^\text{int}_{\mathrm{BB}}(r,\tau,\mu)=
\iint_0^\tau~
\frac{
\text{Tr}\left[
e^{-H(\beta-\tau)}
U^{\dagger}(\vec x,\vec y)
e^{-H(\tau-\tau_1)}
g(\mu)B_{i}(\vec x)
e^{-H(\tau_1-\tau_2)}
g(\mu)B_{i}(\vec y)
e^{-H\tau_2}
U(\vec x,\vec y)
\right]
}{
\text{Tr}\left[
e^{-H(\beta-\tau)}
U^{\dagger}(\vec x,\vec y)
e^{-H\tau}
U(\vec x,\vec y)
\right]
}~.
\end{equation}
Here the trace should be understood as a sum over all physical states of the QCD 
Hamiltonian and $\beta=1/T$. Inserting a complete set of 
eigenstates of the QCD Hamiltonian, labelled by the quantum numbers $l,m,n,k$, and 
performing the time integrations, the above expression can be written as
\begin{equation}
W^\text{int}_{\mathrm{BB}}(r,\tau,\mu)=
\frac{
\sum\limits_{n,m,l,k}
\langle k|U^{\dagger}(\vec x,\vec y)| n\rangle
\langle n|g(\mu)B_{i}(\vec x)|l \rangle
\langle l|g(\mu)B_{i}(\vec y)|m \rangle
\langle m|U(\vec x,\vec y)|k\rangle
\frac{e^{-E_j\tau}-e^{-E_n\tau}}
{(E_j-E_n)(E_l-E_m)}
\left[\delta_{jl}
-\delta_{jm}
\right] e^{-E_k(\beta-\tau)}
}{
\sum\limits_{n,k}
|\langle n|U(\vec x,\vec y)|k\rangle|^2
~e^{-E_n\tau}e^{-E_k(\beta-\tau)}
}~. \nonumber
\end{equation}
The $U(\vec x,\vec y)|k\rangle$ is a gauge-invariant representation of a $q\bar{q}$ pair 
separated by a distance $r=|\vec x-\vec y|$. Therefore, in order to have a non-zero 
overlap with this state, the energy eigenstates $|m\rangle,|n\rangle$ should also have a 
contribution from a $q\bar{q}$ pair. On the other hand, $B_i(\vec y)|m\rangle$ represents an 
excited state of a $q\bar{q}$ pair due to interaction with gluons, usually referred to as a hybrid 
state. Hence $|l\rangle$ should also transform as a hybrid state in order to have a non-zero overlap 
with $B_i(\vec y)|m\rangle$. The energy levels of such hybrid states $E_l$ are typically  
well above the ground state $|0\rangle$ with an energy $E_0$. In the zero-temperature limit 
$\beta\to\infty$, only the vacuum contribution should survive in the trace. Furthermore, in the 
large Euclidean time limit, $\tau\to\infty$, the dominant contribution arises when $E_m=E_n=E_0$.
Considering the leading $\tau$-dependent term and reminding ourselves that the energies 
are a function of the separation $r$ between the heavy $q\bar{q}$ pair, the expression for the correlator 
simplifies to
\begin{equation}
W^\text{int}_{\mathrm{BB}}(r,\tau,\mu)\overset{\tau\to\infty}{=}2\sum_{l} \langle 0|g(\mu)B_{i}(\vec x)|l \rangle~ \langle l | g(\mu)B_{i}(\vec y)|0 \rangle \left[\frac{\tau}{E_l(r)-E_0(r)}-\frac{1}{\big(E_l(r)-E_0(r)\big)^2}\right]~. \nonumber
\label{eqn:spint}
\end{equation}
The interaction part of the spin-dependent potential then can be easily calculated from this expression,
\begin{equation}
\Phi^{\rm int}\!\left(r=|\vec x-\vec y|\right)
=
-2\,c_B^2(\mu,M)
\sum_l
\frac{
\langle 0|g(\mu)B_i(\vec x)|l\rangle
\langle l|g(\mu)B_i(\vec y)|0\rangle
}{E_l(r)-E_0(r)}.
\end{equation}
At zero temperature, the correlation function describing spin-dependent interaction 
shows a simple linear dependence on time $\tau$ and its coefficient denotes the 
spin-dependent potential $\Phi^{\rm int}\!\left(r=|\vec x-\vec y|\right)$. 
By performing a linear fit of the lattice data as a function of $\tau$, one can 
thus extract this potential at sufficiently large $\tau$. One can further 
show that such a simple linear dependence on $\tau$ can also be obtained for 
the self component of the spin-dependent correlator, which will contribute to the spin-dependent potential as an additive constant at zero temperature.

\subsection{Finite temperature}
\label{sec:SpinPert}
We now study the finite-temperature behaviour of the spin-dependent
potential by performing a leading-order calculation using resummed HTL perturbation 
theory, starting from \cref{eqn:IntSpin,eqn:SelfSpin}. At LO, the thermal expectation 
value of the Wilson loop appearing in the denominator of \cref{eqn:IntSpin,eqn:SelfSpin} 
can be simply written as $W_T(r,\tau)= N_c +O(g^2)$. At this order, the gauge links appearing 
in the numerator can be set to unit matrices. This results in simplifying the spin-dependent 
correlation functions as
\begin{align}
W^{\text{int}}_{\mathrm{BB}}(r,\tau)
= g^2 G(r,\tau)~,~W^{\text{self}}_{\mathrm{BB}}(r,\tau)= g^2G(r=0,\tau),
\label{eqn:BBCorrHTL}
\end{align}
which are written in terms of the correlator
\begin{align}
G(r=\vert \vec x-\vec y\vert\,,\tau)=\frac{1}{N_c}\iint_0^\tau \langle \text{Tr}_c[ B_{i}(\vec x,\tau_1) B_{i}(\vec y,\tau_2)]\rangle_T=\frac{1}{4\,N_c}\epsilon_{imn}\epsilon_{icd} \iint_0^\tau \Big\langle \text{Tr}_c[F_{mn}(\vec x,\tau_1)F_{cd}(\vec y,\tau_2)] \Big\rangle_T 
\label{eqn:HTLSSCorr}
\end{align}
The chromomagnetic fields can be written in terms of the non-Abelian field strength tensor $F_{jk}=\partial_j A_k-\partial_k 
A_j+ig[A_j,A_k],~ i,j,k = 1,2,3$ where $A_j=A_j^a t^a$ is the non-Abelian gauge field that transforms under the 
adjoint representation of the gauge group written in terms of the generators $t^a,~a=1\cdots N_c^2-1$. At LO, 
non-Abelian interactions can be neglected, which leads to
\begin{align}
G(r,\tau)=\frac{\epsilon_{imn}\epsilon_{icd}}{N_c} \iint_0^\tau \Big\langle \partial_m A_n(\vec x,\tau_1)\partial_c A_d(\vec y,\tau_2) \Big\rangle_T \nonumber
\end{align}
The derivatives appearing in the expectation values can be calculated in the Fourier space with four-momentum $p_n\equiv(\vec{p},\omega_n)$, where the Matsubara frequencies $\omega_n= 2n\pi T$ are discrete,
\begin{eqnarray}
G(r,\tau)=-\frac{\epsilon_{imn}\epsilon_{icd} \text{Tr}_c(T_aT_b)}{N_c}  \iint_0^\tau \sumint_p \sumint_k  e^{i(\vec k\cdot \vec y+ \vec p\cdot \vec x)} p_m k_c \langle A_n^a(\vec p)A_d^b(\vec k) \rangle \nonumber
\end{eqnarray}
We denote $\sumint_p\equiv T\sum_{n}\int\frac{d^3\vec{p}}{(2\pi)^3}$. Inserting the HTL propagator, which contains the transverse component of the gluon self-energy $\Pi_T$, one can rewrite 
the above equation as
\begin{equation}
  G(r,\tau)  =C_F \iint_0^\tau ~T\sum_{n=0}^\infty~ e^{i\omega_{n} (\tau_1-\tau_2)}\int \frac{d^3 \vec{p}}{(2\pi)^3} e^{-i\vec{p}\cdot \vec{r}}  \frac{2|\vec{p}|^2 } {\omega_n^2+ |\vec{p}|^2+\Pi_T(\omega_n,\vec p)}~. \nonumber
  \label{2ndstep}
\end{equation}
Performing the integrals over $\tau_1$ and $\tau_2$ for $n=0$ leads to a factor $\tau^2$ whereas for $n\neq 0$ 
gives rise to the second term in the following expression,
\begin{eqnarray}
G(r,\tau)=2C_F \int \frac{d^3 \vec{p}}{(2\pi)^3} e^{-i\vec{p}\cdot \vec{r}}  |\vec{p}|^2 \left[\frac{\tau^2 T }{|\vec{p}|^2+\Pi_T (0,\vec{p})} \right. 
\left. + T~\sum_{n\neq 0} \frac{2-e^{i\omega_n \tau}-e^{-i\omega_n \tau}}{\omega_n^2} \frac{1} {\omega_n^2+|\vec{p}|^2+\Pi_T(\omega_n,\vec{p})} \right]~.
\end{eqnarray}{}
Next, using the spectral representation for the transverse gluon propagator,
\begin{eqnarray}
\frac{1}{\omega_n^2+|\vec{p}|^2+\Pi_T(\omega_n,\vec{p})} = \int_{-\infty}^{\infty} \frac{dq^0}{\pi} \frac{\rho_T(q^0,\vec{p})}{q^0-i \omega_n} \nonumber
\end{eqnarray}
and performing the sum over the Matsubara frequencies, one can calculate the correlator 
$G(r,\tau)$ analytically, we get,
\begin{equation}
\begin{aligned}
G(r,\tau)
=&\;2C_F \int \frac{d^3 \vec{p}}{(2\pi)^3}
e^{-i\vec{p}\cdot \vec{r}} |\vec{p}|^2~
\Bigg[\frac{\tau}{|\vec p|^2+\Pi_T(0,\vec p)}-\int \frac{dp_0}{\pi}
\frac{\rho_T(p_0,\vec p)}{p_0}\frac{
1+e^{\beta p_0}
-e^{(\beta-\tau)p_0}-e^{\tau p_0}
}{p_0 (e^{\beta p_0}-1)}
\Bigg] .
\end{aligned}
\label{eqn:htlst_}
\end{equation}

We see that, spin-dependent correlator have a linear as well periodic dependence on $\tau$. This information 
will be used later to interpret our lattice data. The interaction part of the spin-dependent potential can  then be calculated by performing an analytic continuation of the correlator $G(r,\tau)$ in real 
time using the relation 
\begin{equation}
  \lim_{t\to\infty}\frac{e^{(\beta-it)p_0}-e^{itp_0}}{p_0}=-2\pi i\delta(p_0)  
\end{equation}
such that
\begin{equation}
i\partial_t G(r,\tau\to it)= -2\,C_F  \int \frac{d^3 \vec{p}}{(2\pi)^3} e^{-i\vec{p}\cdot \vec{r}} |\vec{p}|^2 \left[\frac{1 }{|\vec{p}|^2+\Pi_T (0,\vec{p})} +2\,i\,T \lim_{p_0\to0} \frac{\rho_T(p_0,\vec p)}{p_0}\right]~.
\end{equation}
Using \cref{defss}, the spin-dependent potential can be calculated at leading order in 
HTL perturbation theory,
\begin{align}
\nonumber
\Phi^{\text{spin}}(r)&=g^2\lim_{t\to\infty}i\partial_t\left[G(0,\tau \to it)-\frac{4}{3} \vec s_1 \cdot \vec s_2\, G(r,\tau\to it)\right]\\
&=-2\,g^2 C_F  \int \frac{d^3 \vec{p}}{(2\pi)^3} \left (1-\frac{4}{3} \vec{s_1}\cdot\vec{s_2}\,e^{-i\vec{p}\cdot \vec{r}}\right) |\vec{p}|^2 \left[\frac{1 }{|\vec{p}|^2+\Pi_T (0,\vec{p})}
+2\,i\,T \lim_{p_0\to0} \frac{\rho_T(p_0,\vec p)}{p_0}\right]
\label{eq:SpinPotPert}
\end{align}
Similar to the static potential, the spin-dependent potential between a heavy $q\bar{q}$ pair also contains an 
imaginary part at finite temperatures. However, there is an important qualitative difference. Whereas the 
transverse or magnetic gluons contribute to the spin-dependent potential, the static potential arises primarily 
due to longitudinal or electric gluons. Since magnetic gluons interact non-perturbatively at any temperature, the 
spin-dependent correlators at finite temperatures are inherently non-perturbative and therefore the corresponding 
potential can only be calculated using non-perturbative techniques. This fact is already inherent within our 
perturbative calculation, since the self-energy $\Pi_T(0,\vec p)$ of transverse gluons and their spectral function $\rho_T(p_0,\vec p)$ are infrared divergent. Proper regulation of these divergences requires next-to-leading-order expressions for $\Pi_T$ and $\rho_T$ in \cref{eq:SpinPotPert}, which are intrinsically non-perturbative.

Nevertheless, one can obtain estimates of the real and imaginary parts of the spin-dependent
potential within HTL by introducing a non-perturbative mass term $m_T\sim g^2 T/\pi$ in the expressions for
self-energy and the spectral function $\Pi_T$ and $\rho_T$, thereby making them infrared safe. As a result, 
one obtains the following expressions for the real and imaginary parts of the spin-dependent potential,
\begin{align}
\Phi^{\text{spin}}_{\text{re}}(r) &= -\frac{C_Fg^2}{\pi^2} \int_0^\infty dp \frac{p^4}{p^2+m_T^2} \left[1-\frac{4}{3} \vec s_1\cdot\vec s_2\frac{\sin pr}{pr}\right]~, \label{eqn:Realpartpertspinspin} \\
\Phi^{\text{spin}}_{\text{im}}(r) &= \frac{C_Fg^2m_D^2}{4\pi\beta} \int_0^\infty dp \frac{p^3}{(p^2+m_T^2)^2} \left[1-\frac{4}{3}\vec s_1\cdot\vec s_2\frac{\sin pr}{pr} \right]~.\label{eqn:Imagpartpertspinspin}
\end{align}
At large distances $rT \gg 1$, both the real and imaginary parts of the spin-dependent potential 
saturate to a constant value, which is larger in magnitude for the pseudoscalar compared to the vector channel. 
At short distances $rT \ll 1$, however, the potentials in \cref{eqn:Realpartpertspinspin,eqn:Imagpartpertspinspin} 
are not valid. 

One has instead to perform a multipole expansion of the spin-dependent potential within pNRQCD
to understand its properties in the region $rT \ll 1$. After integrating out the gauge fields at the scale $1/r$, 
the relevant degrees of freedom are the color-singlet and octet fields $S(\mathbf r,t)$, $O(\mathbf r,t)$, and 
ultrasoft gluons. For the sake of completeness, we mention here the pNRQCD Lagrangian, which contains the 
singlet and octet propagator terms, along with their interactions with ultrasoft gluon fields,
\begin{align}
\mathcal L_{\rm pNRQCD}
&=
S_1^\dagger
\left[
i\partial_0
-
V_s^{(0)}(r)
+
\frac{3}{4}V_s^{\rm SS}(r)
\right]S_1
+
\mathbf S_3^\dagger\cdot
\left[
\left(
i\partial_0
-
V_s^{(0)}(r)
-
\frac{1}{4}V_s^{\rm SS}(r)
\right)\mathbf 1_3
\right]\mathbf S_3
\nonumber\\
&\quad
+
O_1^\dagger
\left[
iD_0
-
V_o^{(0)}(r)
+
\frac{3}{4}V_o^{\rm SS}(r)
\right]O_1
+
\mathbf O_3^\dagger\cdot
\left[
\left(
iD_0
-
V_o^{(0)}(r)
-
\frac{1}{4}V_o^{\rm SS}(r)
\right)\mathbf 1_3
\right]\mathbf O_3
\nonumber\\
&\quad
+
\frac{V_A(r)}{\sqrt{2N_c}}
\Bigg[
O_1^\dagger\,
\mathbf r\cdot g\mathbf E\,S_1
+
\mathbf O_3^\dagger
\mathbf r\cdot g\mathbf E
\,\mathbf S_3
+
S_1^\dagger\,
\mathbf r\cdot g\mathbf E\,O_1
+
\mathbf S_3^\dagger
\mathbf r\cdot g\mathbf E
\,\mathbf O_3
\Bigg]
\nonumber\\
&\quad
+
\frac{c_B V_A^S(r)}{M\sqrt{2N_c}}
\Bigg[
S_1^\dagger\,g\mathbf B\cdot\mathbf O_3
+
\mathbf S_3^\dagger\cdot g\mathbf B\,O_1
+
O_1^\dagger\,g\mathbf B\cdot\mathbf S_3
+
\mathbf O_3^\dagger\cdot g\mathbf B\,S_1
\Bigg]
+\cdots .
\label{eqn:pnrqcdL}
\end{align}
We include only those terms in the Lagrangian that we require for our discussion.
The fields $S_1(O_1)$ and $\mathbf{S}_3(\mathbf{O}_3)$ are related to the pseudoscalar 
and vector color-singlet (octet) states of a static quark-antiquark pair, respectively. 
The Lagrangian describes the propagation of color-singlet and octet fields and the 
interaction of the later with gluon fields manifest through the covariant derivative.
The third line in \cref{eqn:pnrqcdL} represents the singlet-to-octet transition due to 
dipole interactions caused by $\mathbf r\cdot \mathbf E$, which do not mix the spin-singlet 
and spin-triplet states. A mixing between these spin states can only arise due to the spin-dependent 
interaction~\cite{Yang:2024ejk} written in the fourth line of the Lagrangian. The coefficients 
$V_A, V_A^S$ are the Wilson coefficients that are obtained in this effective theory after integrating 
out the UV modes. Quantities $V_{s,o}^{(0)}(r)$ denote static potentials between the singlet and octet states, 
respectively, and $V_{s,o}^{\rm SS}(r)$ denote the spin-dependent contribution to corresponding 
static potentials at zero temperature. The terms in the third line of \cref{eqn:pnrqcdL} will contribute 
to the static potential, which has been derived earlier~\cite{Brambilla:2008cx} to be,
\begin{equation}
\delta V(r)
=-i\frac{V_A^2(r)}{2N_c}\,\frac{r^2}{3}
\int_0^\infty dt\, e^{-i\Delta V(r)t}\,
\Big\langle
g E_i^a(t)\,
U_{\mathrm{adj}}^{ab}(t,0)\,
g E_i^b(0)
\Big\rangle_T~,a,b=1,...,N_c^2-1~.
\label{eqn:pnrqcd_static}
\end{equation}
The $U_\text{adj}(t,0)$ represent gauge links in the adjoint representation, introduced in order to implement  
propagation of color octet states in a gauge-invariant manner.  The $\Delta V$ in \cref{eqn:pnrqcd_static} 
represents the difference between the color octet and singlet potentials at zero temperature, which is independent 
of the spin of the states. For recent lattice studies of such a correlator see \cite{Brambilla:2025cqy}.

We have also calculated the correction to the spin-dependent potential arising due to  
spin-changing interactions in \cref{eqn:pnrqcd_static}, whose expression is
\begin{equation}
 \delta V^{\rm spin}(r)
=
-i\frac{c_B^2 \left(V^S_A(r)\right)^2}{4N_c M^2}
\left(
1-\frac{4}{3}\,\mathbf s_1\cdot\mathbf s_2
\right)
\int_0^\infty dt\,
e^{-i\Delta V(r)t}\,
\Big\langle
g B_i^a(t)\,
U_{\rm adj}^{ab}(t,0)\,
g B_i^b(0)
\Big\rangle_T .
\label{eqn:pnrqcd_spin}
\end{equation} 
Comparing the correction terms derived in \cref{eqn:pnrqcd_static} and \cref{eqn:pnrqcd_spin},  
we observe that the static part of the potential is suppressed by a factor $r^2$ compared to its 
spin-dependent counterpart at short distances. Let us now try to understand finite temperature effects 
qualitatively from a LO calculation of \cref{eqn:pnrqcd_spin}. Following the LO calculation 
of \cref{eqn:pnrqcd_static} in Ref.~\cite{Brambilla:2008cx}, the finite temperature correction to the LO static
potential in pNRQCD due to spin interactions can be derived to be
\begin{equation}
\delta V^{\mathrm{spin}}_{T,\mathrm{LO}}(r)
=
\frac{C_F\alpha_s}{M^2}
\left(
1-\frac{4}{3}\,\mathbf{s}_1\cdot\mathbf{s}_2
\right)
\Bigg[
\frac{2}{\pi}T^3
\left(\frac{\Delta V}{T}\right)
f\left(\frac{\Delta V}{T}\right)
-i(\Delta V)^3 n_B(\Delta V)
\Bigg]~.
\label{eqn:pNRQCD_imag}
\end{equation}
Here, $n_B$ denotes the Bose-Einstein distribution function, and $f(z)=\int dx\, \frac{x^3}{e^x-1}\,
\mathcal{P}\frac{1}{x^2-z^2}$, where $\mathcal{P}$ represents the principal value.
At short distance scales, $\Delta V\gg T$, and the quantities 
$f(\Delta V/T)\sim -\frac{\pi^4}{15}\left(\frac{T}{\Delta V}\right)^2$ 
and $n_B(\Delta V)\sim \exp(-\Delta V/T)$. As a result, the real part of the spin-dependent 
correction term is $\delta V^{\rm spin}_{T,\mathrm{LO}}\sim rT^4/M^2$, whereas its imaginary part is 
$\delta V^{\rm spin}_{T,\mathrm{LO}}\sim \alpha_s^4/(r^3M^2)\exp(-\Delta V/T)$. At short distances, the
finite-temperature correction to the real part vanishes linearly with $r$ and thus is
sub-dominant compared to the correction due to spin interactions at zero temperature. 
The imaginary part also vanishes exponentially fast at short distances.

Returning back to the discussion of static potential within pNRQCD, described by 
\cref{eqn:pnrqcd_static}, its real part at finite temperatures $\sim r^3T^4$ is 
also suppressed compared to the zero-temperature case at short distances. Its
imaginary part $\sim \alpha_s^4/r\exp(-\Delta V/T)$ vanishes exponentially 
fast at short distances. Compared to the spin-dependent potential, the 
static contribution has an additional $r^2$ suppression. For order-of-magnitude estimates, 
we compare the imaginary part of the spin-dependent potentials with the corresponding 
static potentials. The ratio between them are $R_{\rm PS}\sim 3/(r M)^2$ and 
$R_{\rm V}\sim 1/(rM)^2$ for the pseudo-scalar and vector channels respectively. This suggest that perturbatively the spin-dependent potential becomes dominating over the static contribution at the scale $r\lesssim 1/M$.

\section{Lattice Implementation}
\label{sec:LatticeDetails}

\subsection{Details of the lattice simulations}

In this work, we focus on calculating the spin-dependent potential in QCD without dynamical fermions, 
whose action is described by the standard Wilson gauge action.
The gauge configurations have been generated using a Monte Carlo algorithm with heat-bath updates 
and 4 over-relaxation steps per update. We perform our study on a Euclidean space-time 
lattice of spatial size of $N_s=68$ and at a fixed temperature $T=1.5~ T_d$, where 
$r_0 T_d = 0.7457(45)$ is the deconfinement temperature in SU(3) expressed in terms 
of the Sommer scale $r_0$~\cite{Francis:2015lha}. Our simulations are carried out on two different choices of 
the temporal extent of our lattice box $N_\tau=16, 20$, where $T=\frac{1}{aN_\tau}$ for performing continuum 
estimates of different physical observables. We set the lattice spacing in physical units using $r_0/a$ from~\cite{Burnier:2017bod} where the Sommer scale is $r_0 = 0.472~\text{fm}$~\cite{Sommer:2014mea}.

In order to regularize the ultraviolet divergences present in the correlators, we perform  
gradient flow on the gauge fields. This procedure introduces a scale at LO perturbation theory 
which is a function of the flow-time ${\tau}_F$, such that fluctuations of the gauge fields at 
length scale $<\sqrt{8{\tau}_F}$ are removed. In our case, we have implemented Zeuthen 
flow~\cite{Ramos:2015baa} on gauge fields, which uses $\mathcal{O}(a^2)$ Symanzik-improved 
gauge action. The flow-time $\tau_F$ is adjusted for different lattice spacings such that $\tau_F$ 
remains constant in physical units.

\begin{table}[H]
    \centering
    \begin{tabular}{|c|c|c|c|c|c|}
    \hline
     $T/T_d$ & $N_\tau$ & $\beta$ & a (fm) & $N_{\text{configs}}$ & $\sqrt{8{\tau}_F}$ (fm)\\
     \hline
     \hline
     & 16 & 6.870 & 0.0263 & 2000 &  \\
     1.5 & & & & & 0.0334, 0.0409, 0.0578 \\
     & 20 & 7.049 & 0.0211 & 2000 & \\
     \hline
    \end{tabular}
    \caption{The parameters used for our lattice computations in quenched SU(3) gauge theory with a spatial extent $N_s=68$.}
    \label{tab:QCDparam}
\end{table}

\subsection{Observables}

In order to extract the self and interaction parts of the spin-dependent potential between a static $q\bar{q}$ pair 
separated by distance $r$, we first calculate the correlator in \cref{eqn:correlatorBB} on a Euclidean  
space-time lattice. The first term is the correlation between two temporal Wilson lines of length $\tau$ 
in the Euclidean time direction and separated in the spatial direction by a distance $r$. The second term
in \cref{eqn:correlatorBB} comprises two terms discussed in \cref{eqn:SelfSpin} and \cref{eqn:IntSpin} 
respectively, with two chromomagnetic field operator insertions at time $\tau_1,\tau_2$ on the same or 
two different Wilson lines. The correlators are calculated in Coulomb gauge, which is implemented 
according to the procedure outlined in Ref.~\cite{Giusti:2001xf}. Extracting the static heavy quark 
potential from the Wilson line correlator in a gauge choice which is a local function of $\tau$ \cite{Philipsen:2002az}, 
e.g., Coulomb gauge, instead from a gauge-invariant Wilson loop is a well-justified procedure at zero 
temperature~\cite{Philipsen:2001ip}. Both these methods lead to the same results for the 
potential but with a better signal-to-noise ratio for the one extracted from Wilson line 
correlators. At finite temperatures, similarity between the extracted static potentials 
from both these methods was shown within perturbation theory in Ref.~\cite{Burnier:2013fca}.

We have implemented the color-magnetic fields on the lattice site $n$ according to the 
clover-improved discretization, which consists of a sum over four neighboring plaquettes 
${U_{jk}(n)}$ along spatial directions $j,k$ with same orientation, 
\begin{align}
a^2{B}_i(n) = a^2 \frac{1}{2}\epsilon_{ijk} g F_{jk}(n) &= K_{jk}(n) - \frac{1}{3} \text{Tr}_c (K_{jk})\mathrm{I},  \label{eqn:CloverImpB} \\
K_{\mu\nu}(n)&=-\frac{i}{8}\Big[Q_{\mu\nu}-Q_{\nu\mu}\Big](n),
~~~~Q_{\mu\nu}(n) = \Big[U_{\mu\nu}+U_{-\nu\mu}
    +U_{-\mu-\nu}+U_{-\nu\mu}\Big](n)~. \nonumber
\end{align}
The correlators defined in \cref{eqn:SelfSpin,eqn:IntSpin} requires integrating 
over the temporal locations of color-magnetic insertions $\tau_1,\tau_2\leq\tau$,  which increases their 
computational cost significantly. 

At finite temperatures we can calculate correlators of temporal extent $\tau \leq \beta$, 
the maximum size allowed along the Euclidean time direction. Since at higher temperatures 
the temporal size of the lattice shrinks, it is not possible to unambiguously extract the 
ground state energy of the correlation function due to contamination from the higher excited 
states.  Hence potentials can not be extracted directly from the Euclidean correlators. 
Instead, we perform an analytic continuation of the Euclidean correlators to real-time 
and then extract the potential in the limit $t\rightarrow \infty$. This will be discussed 
further in Section~\ref{sec:ExtractionPotential}.

\section{Extraction of the spin-dependent potential in lattice QCD}

\subsection{Strategy for extracting the spin-dependent potential}
\label{sec:ExtractionPotential}
Analytic continuation of the correlators from Euclidean to real time 
is not uniquely defined. This is due to the fact that the correlator is only known 
on a discrete number of points on the lattice with a limited precision due to statistical 
uncertainties. One therefore has to use additional physics-motivated inputs for the potential 
in order to reconstruct real-time correlators. However presence of an imaginary part 
in the spin-dependent potential at finite temperatures, as we have seen perturbatively, 
makes the analytic structure of the spin-dependent correlator more complicated than in the zero-temperature case.
This is similar to the problem of the analytic continuation of Wilson loops or Wilson-line correlators used to extract the static thermal potential, which also has an imaginary component.
The extraction of the static potential from Wilson line correlators has been studied extensively on the lattice~\cite{Burnier:2014ssa, Bala:2019cqu, Bala:2020tdt, Ali:2025iux}. In particular, it has been shown in Refs.~\cite{Bala:2019cqu, Ali:2025iux} that the analytic structure of the correlator calculated in leading-order resummed perturbation theory helps us understand the analytic continuation procedure for lattice data. The idea behind this is the decomposition of $\log W_{T}(r,\tau)$ in the following form,
\begin{equation}
\log W_{T}(r, \tau) = -V_{re}(r) \tau + \int_{-\infty}^{\infty} \sigma(r, \omega) \left[ e^{\omega \tau} + e^{\omega (\beta - \tau)} \right] , d\omega .
\label{eqn:W_lp}
\end{equation}
Such an analytic structure is observed in perturbative calculations of the correlator~\cite{Laine:2006ns} and has been verified to hold nonperturbatively~\cite{Bala:2019cqu, Ali:2025iux} over a wide range, $0 \ll \tau \ll \beta$. Perturbatively $\sigma(r,\omega)$ is related to the electric component of the gluon spectral function~\cite{Ali:2025iux}. Using the further constraints on the condition for the existence of the potential from \cref{defpot}, one can derive the following expression,
\begin{align}
W_T(r, \tau) &= A(r) \exp\Bigg[-V_{\text{re}}(r) \tau - \frac{\beta\,V_{\text{im}}(r)}{\pi }\log\left(\sin\left[\frac{\pi\tau}{\beta}\right]\right)+...\Bigg]~.
\label{eqn:Param_WL}
\end{align} 

We use this form to fit the lattice data and extract the static potential from it, as shown in \cref{app1}. 

In the case of spin-dependent correlator, we also observed in \cref{eqn:htlst_} that the correlator consists of a linear and periodic part in $\tau$. This motivates naively to generalize the ansatz in \cref{eqn:W_lp} for the spin-dependent correlator as well, which is given by,
\begin{equation}
 W^{\text{spin}}_{\text{BB}}(r, \tau) = -V^{\text{spin}}_{\text{re}}(r) \tau + \int_{-\infty}^{\infty} \sigma^{\text{spin}}(r, \omega) \left[ e^{\omega \tau} + e^{\omega (\beta - \tau)} \right] \, d\omega~,
\label{eqn:spin_basic}
\end{equation}
which leads to the following parametrization of the spin correlator, 
\begin{align}
W^{\text{spin}}_{\text{BB}}(r, \tau) &= A^{\text{spin}}(r)-V^{\text{spin}}_{\text{re}}(r) \tau - \frac{\beta V^{\text{spin}}_{\text{im}}(r)}{\pi }\log\left(\sin\left(\frac{\pi\tau}{\beta}\right)\right)+....
\label{eqn:Param_spin}
\end{align}

Note that the relation \cref{eqn:spin_basic} does not involve the logarithm of the $W_{\text{BB}}$ 
correlator. We fit our lattice data for the spin correlators using \cref{eqn:Param_spin} with three free parameters in order to extract the real and imaginary parts of the spin-dependent potential. Indeed, we find that this fit parametrization describes the lattice data with a very good $\chi^2/{\rm dof}$. A similar form to \cref{eqn:Param_spin} has also been used to extract the correction to thermal potential due to non-zero density \cite{Goswami:2026ixn}. 
 To obtain the physical spin correlators, we first need to perform a continuum extrapolation of correlators at a fixed 
flow-time, followed by a zero flow-time extrapolation and then a proper renormalization of the correlators. 
These steps are detailed in the following sections.

\subsection{Continuum estimates of the spin-dependent correlators}

The spin-dependent correlator calculated on the lattice needs to be renormalized 
due to the presence of chromomagnetic field operators. We use the gradient-flow 
technique for this purpose. The gradient flow suppresses ultraviolet fluctuations 
of the gauge fields which exist above a characteristic scale $\mu_F \sim 1/\sqrt{8\tau_F}$, 
and allows us to define the correlator at the flow scale $\mu_F$.

We then perform a continuum estimation of the renormalized integrand appearing in
\cref{eqn:SelfSpin,eqn:IntSpin} calculated at the scale $\tau_F$, which we denote by
$W^{\text{int/self}}_{\text{BB}}(r,\tau,\tau_1,\tau_2,\tau_F)$. The continuum estimation
at fixed physical values of $rT$, $\tau T$, $\tau_1T$, $\tau_2T$, and $\tau_FT^2$ 
is carried out using two lattice spacings listed in \cref{tab:QCDparam}. However, 
a smooth interpolation of the data for the coarser lattice needs to be performed in 
order to have the correlator at all points where data from the finer lattice is available. 
The lattice correlator decreases rapidly with separation $|\tau_1-\tau_2|$. Instead of 
performing an interpolation of the correlator data itself, we interpolate between the data 
for the ratio
\begin{equation}
     R^{\text{int/self}}_{\text{BB}}
     (rT,\tau T,\tau_1T,\tau_2T,\tau_FT^2)\big|_{\text{lat}}
     =
     g^2 C_F\frac{
     W^{\text{int/self}}_{\text{BB}}
     (rT,\tau T, \tau_1T,\tau_2T,\tau_F T^2)\big|_{\text{lat}}
     }{
     W^{\rm LO}_{\text{BB}}
     (rT,\tau_1 T,\tau_2 T,\tau_F T^2)\big|_{\text{lat}}
     } .
\label{eqn:ratio_spin}
\end{equation}
where the tree-level LO correlator can be analytically calculated \cite{Stendebach:2022} in the 
Fourier space giving us
\begin{align}
W^{\rm LO}_{\rm BB}(r,\tau_1,\tau_2,\tau_F;N_\tau)\big|_{\rm lat} &=
g^2 C_F \int_{-\pi}^{\pi} \frac{d^3q}{(2\pi)^3} e^{iq_z r} \left(\sum_i c_i^2\sum_i s_i^2 -\sum_i c_i^2 s_i^2\right) \left[\mathcal{K}(\hat{\mathbf q}^{\,2})-\mathcal{L}^{-1}\left(\frac{\mathcal{K}(s/2+\hat{\mathbf q}^{\,2})}{s}\right)_{\tau_F}\right]
\label{latt_free} \\
&\mathcal{K}(x) = \frac{e^{\bar{q}N_\tau(1-\Delta\tau T)}+e^{\bar{q}N_\tau \Delta\tau T}}{(e^{\bar{q}N_\tau}-1) \sinh{\bar{q}}},~~~ \bar{q}=2\sinh^{-1}{\left(\frac{\sqrt{x}}{2}\right)}\nonumber.
\end{align}
The quantities appearing in the free correlator are defined 
as $\hat\omega_n=2\sin(\omega_n/2)$, $\hat q_i=2\sin(q_i/2)$, 
$s_i=\sin q_i$, $c_i=\cos(q_i/2)$, $\hat{\mathbf q}^{\,2}=\sum_i\hat q_i^2$ and 
$\mathcal{L}^{-1}$ is the inverse Laplace transform calculated at flow-time $\tau_F$.
Dividing by the LO lattice correlator removes the dominant dependence on 
$\Delta\tau\equiv|\tau_1-\tau_2|$ and also reduces the leading tree-level 
cutoff effects \cite{Meyer:2009vj} in the lattice correlator. 
We use multidimensional RBF interpolation with a cubic basis function for the 
$N_\tau=16$ data to obtain the correlator values at the same points, defined 
in terms of the four parameters, $rT$, $\tau_1 T$, $\tau_2 T$, and $\tau T$, 
where measurements on the finer $N_\tau=20$ lattice have been performed. The 
continuum estimates are then obtained by performing a fit to the data for 
$R^{\text{int/self}}_{\text{BB}}$ using the following ansatz, since both the action 
and the correlator have cutoff effects of order $O(a^2)$,
\begin{eqnarray}
 R^{\text{int/self}}_{\text{BB}}
 (rT,\tau T,\tau_1T,\tau_2T,\tau_FT^2)\big|_{\text{lat}}
 =
 R^{\text{int/self}}_{\text{BB}}
 (rT,\tau T,\tau_1T,\tau_2T,\tau_FT^2)\big|_{\text{cont}}
 +\frac{c}{N_{\tau}^2}.
\end{eqnarray}

\begin{figure}[tbp]
    \centering
    \includegraphics[width=0.47\textwidth]{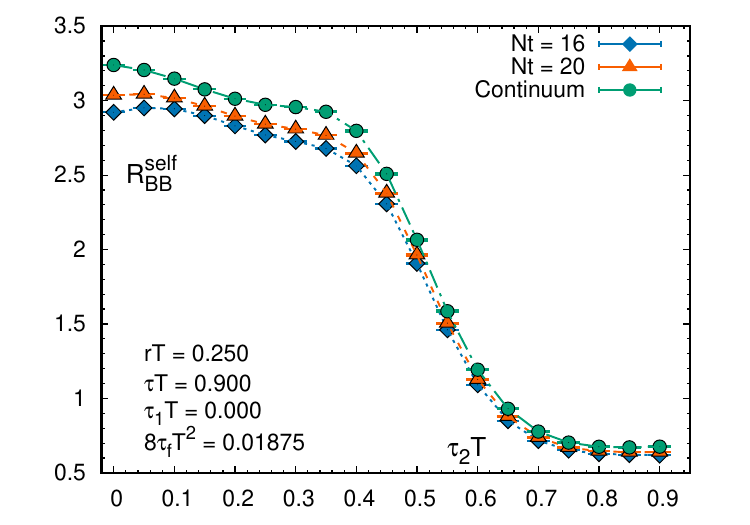}
   \includegraphics[width=0.47\textwidth]{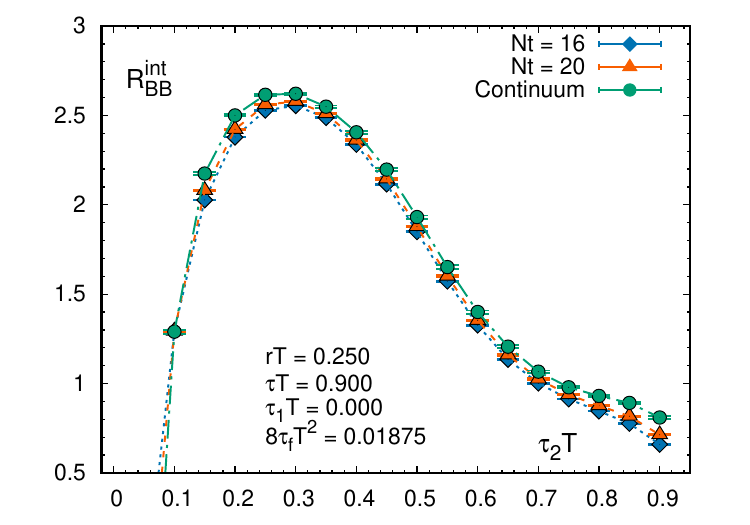}
   \caption{The continuum estimated self (left panel) and interaction part (right panel) of the ratio given in \cref{eqn:ratio_spin}, shown as a function of $\tau_2 T$, where the chromomagnetic insertions are at $\tau_1$ and $\tau_2$, respectively. The total temporal extent of the Wilson lines are $\tau T=0.9$ and the correlators are separated by a distance $rT=0.25$. The values of these correlators for two different lattice cut-offs, corresponding to $N_\tau=16, 20$ respectively are also shown. Results are shown at a finite flow time $8\tau_f T^2=0.01875$. }
    \label{fig:latt_corr_cutoff_effects}
\end{figure}

The data for the ratio $R^{\text{int/self}}_{\text{BB}}$  for two different 
lattice spacings, as well as the continuum-estimated values are shown in 
\cref{fig:latt_corr_cutoff_effects} as a function of $\tau_2T$, keeping the 
other parameters fixed at $rT=0.25, \tau T=0.9, \tau_1T=0, 8\tau_FT^2=0.01875$. 
For the self-spin contribution to the potential shown in the left panel of 
\cref{fig:latt_corr_cutoff_effects}, the lattice correlator at LO perturbation 
is known to be independent of $r$ and decays exponentially as a function of 
$|\tau_1-\tau_2|$, hence our procedure works quite well. For $R^{\text{int}}_{\text{BB}}$, 
the continuum estimation must be performed carefully, since this quantity is not defined 
at the zeros of the LO correlator as a function of $\tau_1-\tau_2$. Consequently, the ratios 
blow up near these zeros, leading to an unstable interpolation. We address this issue by first 
interpolating the ratio of the interaction part of the spin correlator to the LO correlator evaluated at $rT=0$. We 
then multiply this interpolated ratio by $W_{\text{BB}}^{\text{LO}}(rT=0)/W_{\text{BB}}^{\text{LO}}(rT)$ 
to obtain the tree-level improved correlator. Our procedure is succinctly summarized in the following equation,
\begin{equation}
     R^{\text{int}}_{\text{BB}}
     (rT,\tau T,\tau_1T,\tau_2T,\tau_FT^2)\big|_{\text{lat}}
     =
     g^2 C_F\left.
     \frac{
     W^{\text{int}}_{\text{BB}}
     (rT,\tau T,\tau_1T,\tau_2T,\tau_F T^2)\big|_{\text{lat}}
     }{
     W^{\text{LO}}_{\text{BB}}
     (rT=0,\tau_1 T,\tau_2 T,\tau_F T^2)\big|_{\text{lat}}
     }
     \right|_{\text{intp}}\times
     \frac{
     W^{\text{LO}}_{\text{BB}}
     (rT=0,\tau_1 T,\tau_2 T,\tau_F T^2)\big|_{\text{lat}}
     }{
     W^{\text{LO}}_{\text{BB}}
     (rT,\tau_1 T,\tau_2 T,\tau_F T^2)\big|_{\text{lat}}
     }.
     \label{eq:RBB}
\end{equation}

This procedure leads to a reliable interpolation of the lattice data and subsequently allows the 
continuum estimation to be performed. The continuum-estimated ratio $R^{\text{int}}_{\text{BB}}$, 
as a function of $\tau_2 T$, is shown in the right panel of \cref{fig:latt_corr_cutoff_effects}.

\subsection{Zero flow-time extrapolation} \label{RenormBB}

After performing continuum estimation of the ratios $R^{\text{int/self}}_{\text{BB}}$, 
at a flow scale $\tau_FT^2$ we next calculate the renormalized spin-dependent correlator in the 
$\overline{\rm MS}$ scheme at the scale $\bar\mu_b = M_b$. The procedure for relating the 
correlators calculated in the gradient flow and $\overline{\rm MS}$ schemes has been 
outlined in Ref.~\cite{delaCruz:2024cix}. For the sake of completeness, we briefly 
summarize the steps. First, the following NLO relation is used for matching the correlators 
calculated in the gradient-flow scheme to the $\overline{\rm MS}$ scheme \cite{Brambilla:2023vwm, delaCruz:2024cix},
\begin{equation}
W^{\rm flow,\mu_F}_{\text{BB}}(rT,\tau T,\tau_1 T,\tau_2 T)_{\rm self/int}
=
W^{\overline{\rm MS},\bar{\mu}}_{\text{BB}}(rT,\tau T,\tau_1 T,\tau_2 T)_{\rm self/int}
\left[
1+\gamma_0 g^2
\left(
\ln\frac{\bar{\mu}^2}{4\mu_F^2}
+\gamma_E
\right)
\right].
\label{eqn:renor}
\end{equation}
Here $\gamma_0=3/8\pi^2$ is the anomalous dimension at LO. It should be noted that this matching procedure is valid when $|\tau_1 T-\tau_2T| \gg \sqrt{\tau_F T^2}$ 
thus limiting its applicability for very small separations. 
Furthermore, in order to avoid large logarithmic corrections, we perform a matching at the 
minimum-sensitivity scale $\bar{\mu}_{\tau_F}= 1.5\,\mu_F$, and then perform 
a renormalization-group evolution of the correlator at the $\overline{\rm MS}$ scale $\bar{\mu}_b$. This results in the following renormalization coefficient
\begin{eqnarray}
    \ln{Z}_\text{ren}(\bar{\mu}_b,\bar{\mu}_{\tau_F},\mu_F)
    &=&
    \int^{\bar{\mu}_{b}^2}_{\bar{\mu}_{\tau_F}^2}
    \gamma_0 g^2(\bar{\mu}) \frac{d\bar{\mu}^2}{\bar{\mu}^2}
    - \gamma_0 g^2(\bar{\mu}_{\tau_F})
    \left[
    \ln{\left(\frac{\bar{\mu}_{\tau_F}^2}{4\mu_F^2}\right)}
    +\gamma_E
    \right]~.
\end{eqnarray}
We choose the bottom quark pole mass to be $\bar{\mu}_b=4.78\,{\rm GeV}$ \cite{ParticleDataGroup:2024cfk}. 
The RG running of the coupling is performed using a combination of the non-perturbatively determined coupling 
from Ref.~\cite{delaCruz:2024cix} together with the 5-loop beta function. This gives rise to the 
renormalization coefficients $Z_{\rm ren} \simeq  \{1.08,\; 1.06,\; 1.01\}$ for the three flow times 
listed in \cref{tab:QCDparam}, in increasing order of magnitude. The renormalization coefficient 
removes the logarithmic contribution in the correlators arising due to the choice of the flow-time scale.

\begin{figure}[tbp]
    \centering
    \includegraphics[width=0.45\textwidth]{ 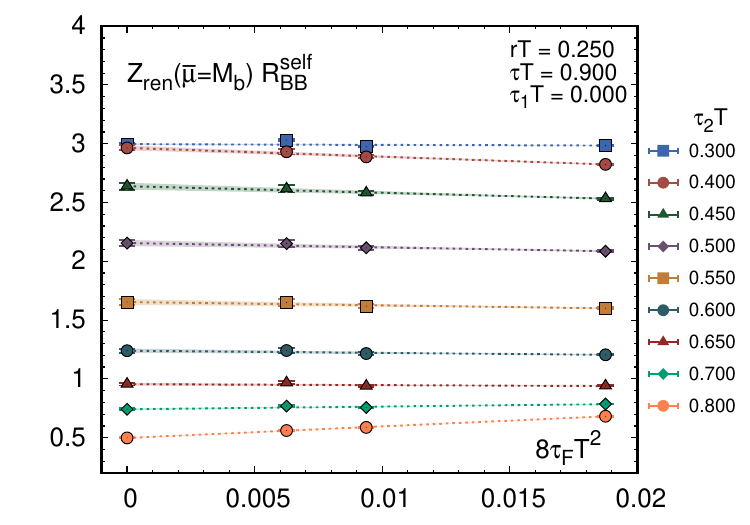}
    \includegraphics[width=0.45\textwidth]{ 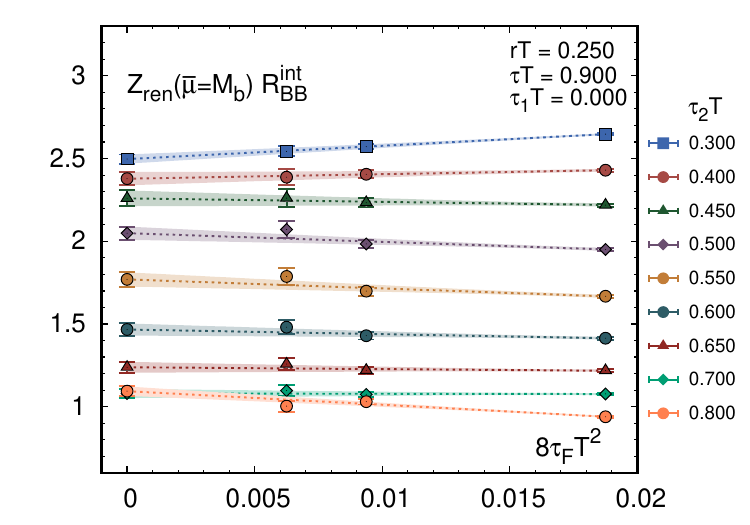}
    \caption{The zero flow-time extrapolation of the self (left panel) and the interaction (right panel) part of the renormalized ratios given in \cref{eqn:ratio_spin}, where the chromomagnetic insertions are at Euclidean times $\tau_2$ and $\tau_1$, respectively. The renormalization is performed at a scale $\bar \mu=M_b$. }
    \label{fig:zero_flow_corr}
\end{figure}{}

\begin{figure}[tbp]
    \centering
    \includegraphics[width=0.45\textwidth]{ 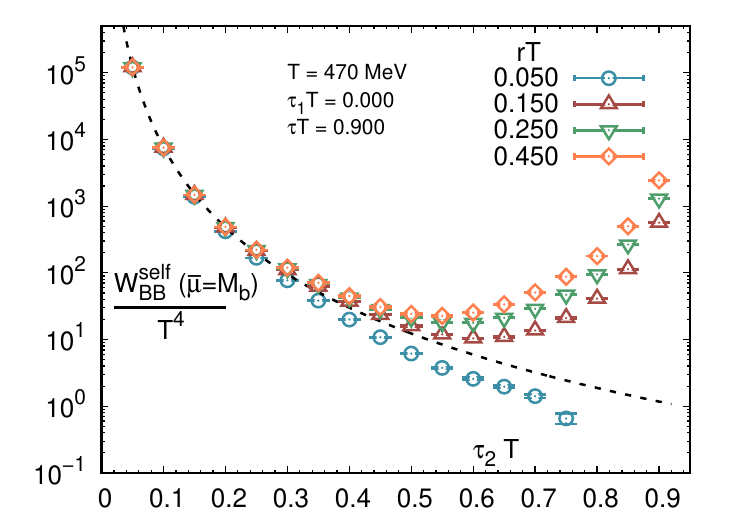}
    \includegraphics[width=0.45\textwidth]{ 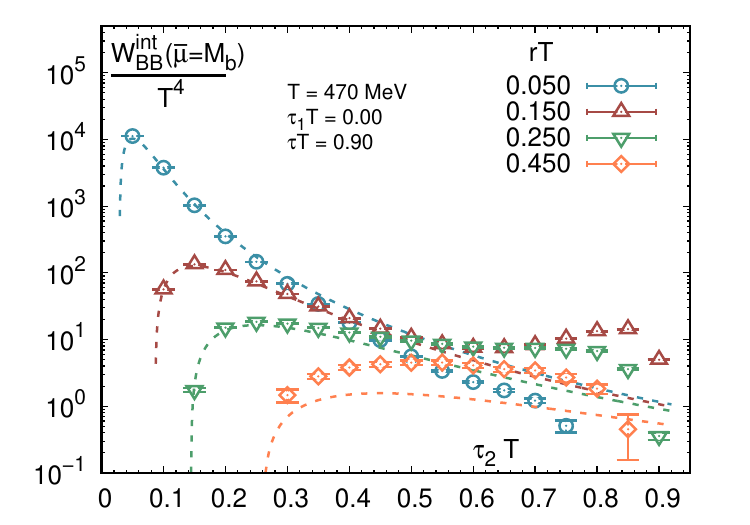}
    \caption{The self (left panel) and the interaction (right panel) part of the zero flow time extrapolated renormalized spin-dependent correlators, which are calculated in the $\overline{\text{MS}}$ scheme at the scale $\bar{\mu} = M_b$, shown as function of $\tau_2$. Our results are compared with the perturbatively determined zero temperature values of the same correlators which are shown as dotted lines.}
    \label{fig:renormalized correlator}
\end{figure}

The resulting data for the renormalized ratios are illustrated in \cref{fig:zero_flow_corr} for 
representative values of the parameters $rT=0.25$, $\tau T=0.9$, $\tau_1T=0$. We observe that each 
renormalized ratio exhibits only a mild dependence on the 
flow-time scale, confirming the validity of the renormalization procedure. To remove gradient-flow artifacts, 
the renormalized values of the spin-dependent correlator ratios in the $\overline{\rm MS}$ scheme are extrapolated 
to zero flow-time using the linear ansatz,
\begin{eqnarray}
 R^{\overline{\rm MS}}_{\rm BB}
 (rT,\tau T,\tau_1T,\tau_2T,\tau_F T^2)\big|_{\rm cont}
 &=&
 R^{\overline{\rm MS}}_{\rm BB}
 (rT,\tau T,\tau_1T,\tau_2T,0)\big|_{\rm cont}
 + c_1\,\tau_F T^2 .
 \label{zfe1}
\end{eqnarray}
Finally the 
spin-dependent correlators at the $\overline{\rm MS}$ scale $\bar\mu_b$ can be 
obtained from the renormalized spin-dependent correlator ratios by multiplying 
with the LO correlators at zero flow time, which can be obtained from,
\cref{latt_free},
\begin{equation}
\frac{W^{\rm LO}_{\text{BB}}(rT,\tau_1 T, \tau_2 T)}{T^4}
=-g^2 C_F\frac{4\pi}{rT} \mathrm{Re} \left [\frac{z(1+z)}{(1-z)^3}\right],
\qquad \text{where}~~
z=\exp\left[-2\pi T (r-i |\tau_1-\tau_2|)\right].
\label{eqn:free_T}
\end{equation}
The results for the self and interaction parts of the spin-dependent correlators are summarized in the left and right panels of 
\cref{fig:renormalized correlator}, respectively.  At small separations, $\tau_2 T \sim 0.1$--$0.15$, 
the self part of the spin-dependent correlator matches with its LO perturbative estimates at $T=0$ shown by a dashed line, independent 
of the spatial separation $r$ between the $q\bar q$ pair. This is expected since at vanishingly small separation 
between the two chromomagnetic field insertions ultraviolet fluctuations are dominantly large.  At larger separations 
between the field insertions, however, noticeable thermal effects are visible in the correlator, which becomes more 
prominent as the separation between the $q\bar{q}$ pair increases. For the interaction part of the spin-dependent correlator arising from chromomagnetic insertions on different Wilson lines, thermal effects are visible over the entire range of $\tau_2 T$ at large separation distances $rT$ between a heavy $q\bar q$ pair. However, at smaller values of $rT$, the dependence for $\tau_2 T<0.3$ can be explained in terms of the corresponding perturbative zero-temperature LO correlator.

{}

\subsection{Integrating over the temporal insertion points of chromomagnetic fields }

Having obtained the spin-dependent correlators in the  $\overline{\mathrm{MS}}$ scheme for different values 
of $\tau_1 T$ and $\tau_2 T$, the next step is to integrate out these coordinates as defined in 
\cref{eqn:SelfSpin,eqn:IntSpin}. However, while performing the extrapolation of the correlator values 
to zero flow time, we have already mentioned that the renormalization procedure is well-defined only for 
$\vert \tau_1T-\tau_2T \vert \gg \sqrt{\tau_F T^2}$, since \cref{eqn:renor} is not valid for smaller separations. 
Therefore, while integrating, we do not include insertion points with separations less than a resolution scale 
$\epsilon$. The scale $\epsilon$ is chosen to be sufficiently small, but still larger than $\sim \sqrt{\tau_F T^2}$. 
This ensures that, while some short-distance ultraviolet fluctuations are regulated by the resolution scale $\epsilon$, 
the desired thermal contributions remain unchanged. In order to see how $\epsilon$ can affect the spin-dependent 
potential, we perform a perturbative calculation of the spin-dependent correlator within HTL perturbation theory in the 
presence of a finite resolution scale $\epsilon$. The resulting integrated correlator can be calculated from 
\cref{eqn:HTLSSCorr}, similarly to the procedure described in \cref{sec:ExtractionPotential}, giving us the following 
expression analogous to \cref{eqn:htlst_}:

\begin{equation}
\begin{aligned}
G(r,\tau,\epsilon)
=&\;2C_F \int \frac{d^3 \vec{p}}{(2\pi)^3}
e^{-i\vec{p}\cdot \vec{r}} |\vec{p}|^2~
 \int \frac{dp_0}{\pi}
\frac{\rho_T(p_0,\vec p)}{p_0}
\Bigg[
(\tau-\epsilon)
\frac{e^{p_0(\beta-\epsilon)}-e^{p_0\epsilon}}
{e^{p_0\beta}-1}
-
\frac{
e^{p_0\epsilon}+e^{p_0(\beta-\epsilon)}
-e^{p_0(\beta-\tau)}-e^{p_0\tau}
}{p_0 (e^{p_0\beta}-1)}
\Bigg] .
\end{aligned}
\label{eqn:htlst_eps}
\end{equation}
As in \cref{sec:ExtractionPotential}, the coefficient of the term which is linearly proportional to $\tau$ gives 
us the real part of the potential, which can now be seen to depend on the resolution $\epsilon$. Similarly, 
the imaginary part is related to the periodic function in $\tau$ and can be easily seen to be independent 
of the resolution scale. In order to understand the effects due to the resolution scale we now explicitly calculate the 
spin-dependent potential at tree level, inserting the following spectral function 
\begin{equation}
\rho_T(p_0,p)=\frac{\pi}{2|\vec p|}(\delta(p_0-|\vec p|)-\delta(p_0+|\vec p|)).
\end{equation}
The interaction part of the spin-dependent potential turns out to be
\begin{equation}
\frac{\Phi_\epsilon^{\text{int}}(rT)}{T^3}
= -g^2 C_F\,\frac{2}{rT}\,\frac{\sinh(2\pi rT)\,\sin(2\pi \epsilon T)}{
\left[\cosh(2\pi rT)-\cos(2\pi \epsilon T)\right]^2} \overset{T \to 0}{=}
-\frac{2g^2 C_F}{\pi^2}
\frac{\epsilon }{\left[r^2+\epsilon^2\right]^2}-\frac{2\pi^2}{15}\,g^2 C_F\,\epsilon T+\mathcal{O}(\epsilon^3T^3).
\label{eqn:ftdelta}
\end{equation}
The first term gives the expected $\delta^3(\vec r)$ potential in the limit when $\epsilon\to 0$, whereas 
all small but finite temperature-dependent terms vanish. The self-spin interaction potential can be 
similarly derived from the above equation at $r=0$,
\begin{equation}
\Phi^{\text{self}}_\epsilon(T)
=-2\pi g^2C_F \cot{(\pi \epsilon T )} \operatorname{cosec}^2{(\pi \epsilon T )}\overset{T \to 0}=-\frac{2 g^2 C_F}{\pi^2}\frac{1}{\epsilon^3}
-\frac{2\pi^2}{15} g^2 C_F\,\epsilon T^4
+T^3\mathcal{O}\!\left((\epsilon T)^3\right).
\label{eqn:divergence}
\end{equation}

This self part of the spin-dependent potential thus contains the same LO short-distance divergence as in 
the case at zero temperature. In addition, it contains regular terms in powers of $\epsilon T$ 
which are artifacts of finite resolution scale and must also be removed while extracting the 
physical potential.

\begin{figure}[tbp]
    \centering
    \includegraphics[width=0.45\textwidth]{ 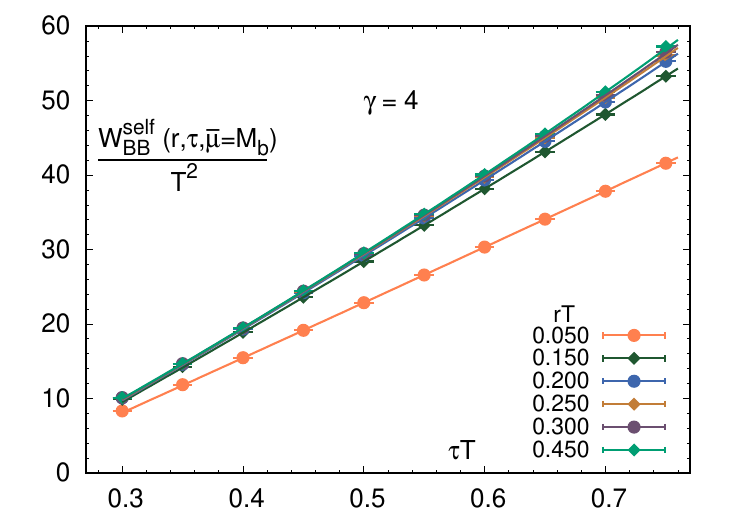}
    \includegraphics[width=0.45\textwidth]{ 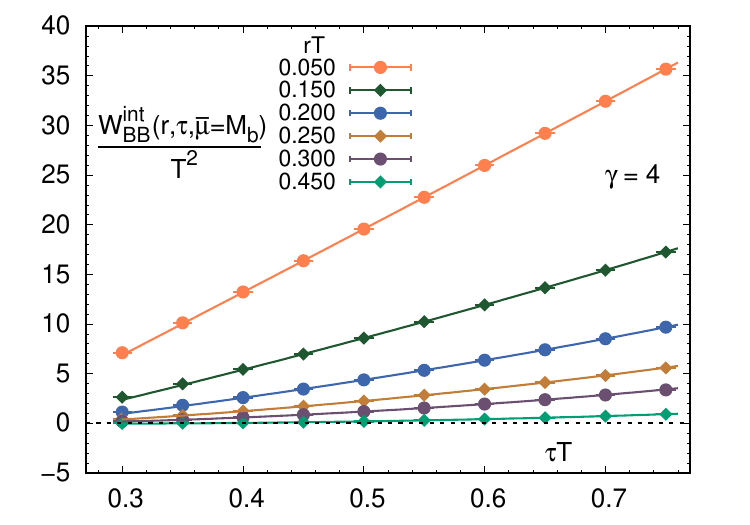}
    \caption{Integrated correlators obtained after performing the integrations, with chromomagnetic insertions on the same Wilson line (left panel) or on two different Wilson lines (right panel), plotted for various values of $rT$. The solid lines denote the fits to the lattice data according to the parametrization in \cref{eqn:Param_spin}. The correlators are calculated at the renormalization scale $\bar\mu=M_b$.}
    \label{fig:int_corr}
\end{figure}

Having understood that, we now perform a numerical integration over the coordinates $\tau_1,\tau_2$ by choosing 
$\epsilon =\gamma \Delta\tau$, for fixed time step $\Delta \tau \,T=0.05$ and different choices 
of $\gamma=4,5,6$. This ensures that the criterion $|\tau_1T-\tau_2T|\gg\sqrt{\tau_F T^2}$ 
is respected. The integrated correlators for $\gamma=4$ are shown in \cref{fig:int_corr}, 
whose slope with respect to $\tau T$ is proportional to the corresponding real part of the 
potential. We next perform a fit to the integrated correlator using the parametrization described 
in \cref{eqn:Param_spin}. 
The fit has been performed over various ranges, with the minimum number of data points fixed to be $5$. 
The results obtained from these various fit ranges are accepted if $\chi^2/\mathrm{dof}\lesssim 2.5$.
The resulting potentials obtained from these different fit ranges yield very stable values for both the 
real and imaginary parts. The final potential is obtained by taking the median over the range $\tau T$ that 
gives a good fit, and the errors are estimated from the $16^\text{th}$ to $84^\text{th}$ percentile interval. 
Some representative fits are also shown in \cref{fig:int_corr} for both the integrated self and interaction part of spin-dependent correlators. 

\begin{figure}[tbp]
    \centering
    \includegraphics[width=0.45\textwidth]{ 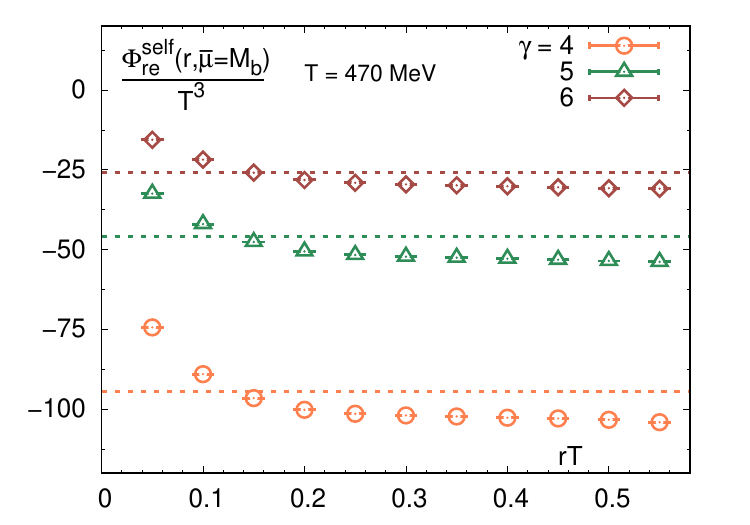}
    \includegraphics[width=0.45\textwidth]{ 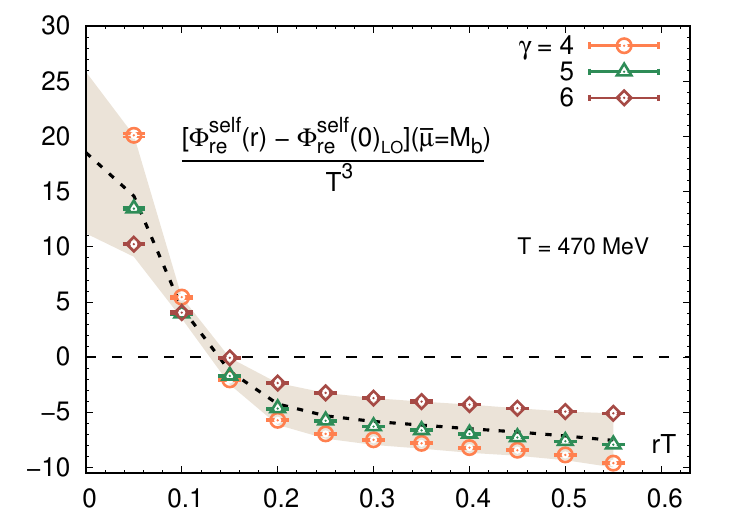}
    \caption{Real self part of the renormalized spin-dependent potential before (left panel) and after 
    (right panel) subtracting the leading divergent term, shown as a function of $rT$, at a temperature $T=470$ MeV.}
    \label{fig:spin_potential_self}
\end{figure}

The real part of the self contribution to the spin-dependent potential calculated from these fits is 
shown in the left panel of \cref{fig:spin_potential_self}.  We observe a large difference between the values 
for different choices of $\gamma$, which can be inferred to arise from the $1/\epsilon^3$ divergence term, as 
derived in \cref{eqn:divergence}. In principle, one should calculate the values of the potential obtained from a 
zero-temperature lattice simulation using the same lattice parameters and subtract their contribution from the finite 
temperature data. However, this is computationally expensive. 
Instead, we subtract the finite-temperature tree-level divergent contribution estimated in \cref{eqn:divergence}, 
with the expectation that this removes the dominant divergent part. To account for discretization effects in the 
numerical integration, we first integrate the tree-level expression in \cref{eqn:free_T} using the same grid 
spacing, $\Delta \tau T=0.05$, as that of the continuum-estimated lattice correlator. The corresponding spin-dependent potential at tree level is then extracted using the 
deriviative at the midpoint. The results for the 
self part of the spin-dependent potential, after subtracting this LO contribution, are shown in the right panel of 
\cref{fig:spin_potential_self}. At large distances $rT\gtrsim0.2$, results for different choices of $\gamma$ are very 
close to each other, indicating the expected cancellation of the divergent term. However, at short distances the LO 
subtracted potential has noticeable dependence on the resolution scale $\gamma$. This arises due to hard gluon 
exchanges between the heavy $q\bar{q}$ pair. Quantifying this effect would require a full NLO calculation, which is beyond the 
scope of the present work. Henceforth we will discuss the physical consequences of the spin-dependent potential 
only at length-scales $rT\gtrsim 0.2$.

\begin{figure}[b]
    \centering
    \includegraphics[width=0.45\textwidth]{ 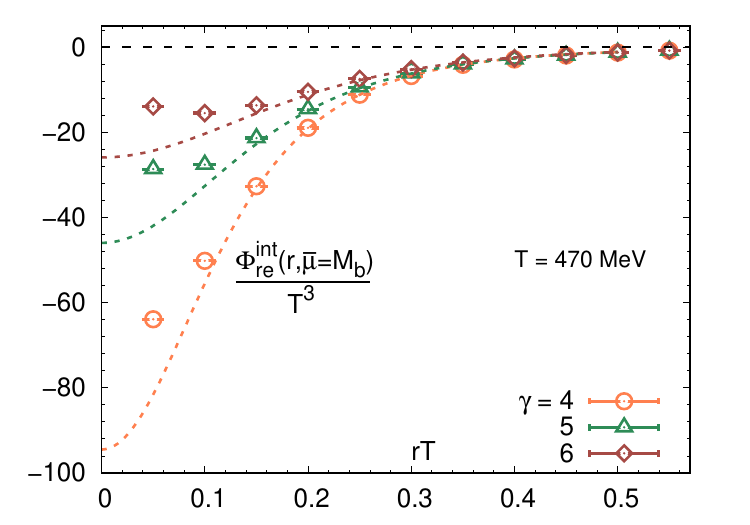}
    \caption{Interaction part of the renormalized spin-dependent potential shown as a function of $rT$, for different choices of the resolution scale $\gamma$. Our lattice results are compared with the results obtained at tree-level, shown as dotted lines.}
    \label{fig:spin_interacting}
\end{figure}

The interaction part of the spin-dependent potential is shown in \cref{fig:spin_interacting}. We again observe 
significant dependence on the choice of the regulator at short distances. In the same figure, we also plot the 
corresponding tree-level potential from \cref{eqn:free_T}. For $rT\gtrsim 0.15$, the potential can thus be
understood in terms of a smeared delta function. However, understanding the deviation from tree-level expectations 
at short distances $rT<0.1$  would require an NLO calculation, since hard gluon exchanges between the heavy $q\bar{q}$ pair again become important. 

\begin{figure}[tbp]
    \centering
    \includegraphics[width=0.45\textwidth]{ 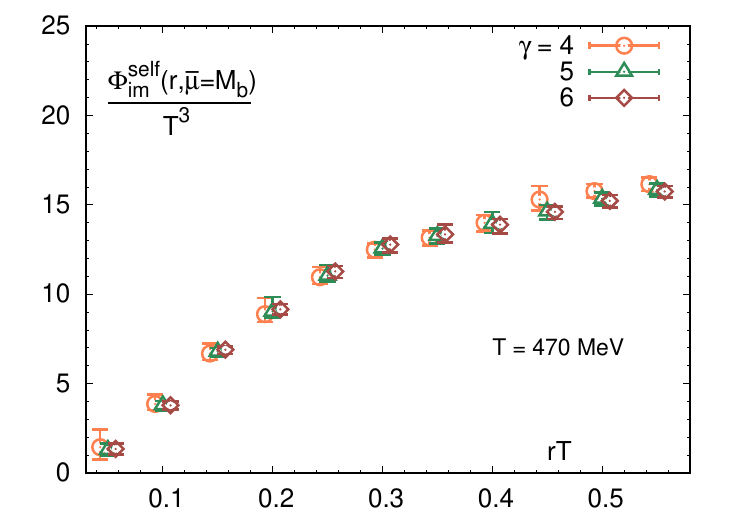}
    \includegraphics[width=0.45\textwidth]{ 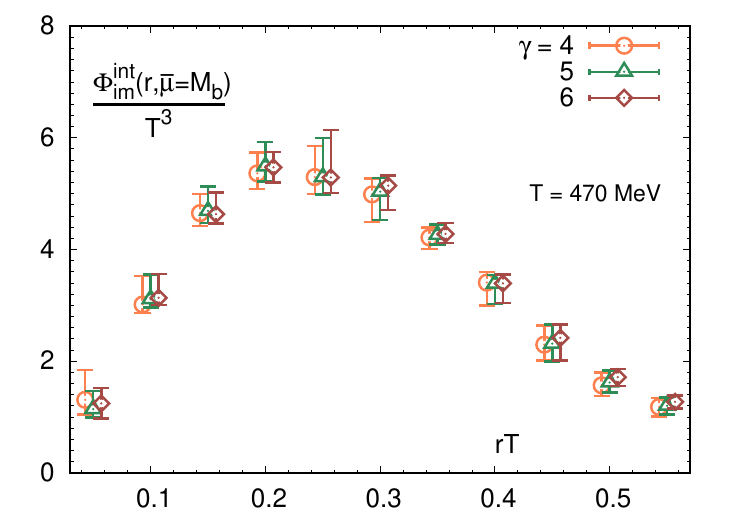}
    \caption{Imaginary part of the self (left panel) and interaction (right panel) part of renormalized spin-dependent potential shown as a function of $rT$ respectively. }
    \label{fig:spin_potential_imaginary}
\end{figure}

The results for the imaginary parts of the self and interaction part of spin-dependent potentials are shown in 
\cref{fig:spin_potential_imaginary}. Unlike their real counterparts, the imaginary contributions are found to be 
insensitive to the choice of regularization parameter $\epsilon$, or equivalently $\gamma$, as also expected from 
the LO expression calculated at a finite resolution. This makes the extraction of the imaginary component of the 
spin-dependent potential much more robust. For the self part, the imaginary part starts from a finite 
value at short distances and saturates to a constant value at long distances. In contrast, the interaction part is similar 
in magnitude at short distances, increases to reach a maximum around $rT \sim 0.2$ and then decreases, eventually 
approaching zero at large distances.

\section{spin-dependent potential \& its impact on quarkonium spectral functions}

\subsection{Spin-dependent potential for charm and bottom quarks}
In this section, we discuss the physical spin-dependent potential
$V(r)$ for quarkonium states, obtained from the
$\overline{\mathrm{MS}}$ potential $\Phi^{\text{int/self}}(r,\bar\mu)$ at the scale
$\bar\mu_b=M_b$ in the previous section. 
To obtain the physical spin-dependent potential, one has
to perform a RG evolution from the $\overline{\mathrm{MS}}$ scale 
$\bar{\mu}_b$ to the scale set by the desired heavy-quark mass $M$,
and subsequently multiply the result by the Wilson coefficient
$c_B^2(\bar{\mu}=M,M)$, which at one-loop order is given by
\cite{Laine:2021uzs},
\begin{align}
    c^2_B(\bar\mu,M)
    =
    1
    +
    2\gamma_0
    \left[
    \ln{\left(\frac{\bar{\mu}^2}{M^2}\right)}
    +2\frac{C_F}{C_A}
    +2
    \right]
    \frac{g^2(\bar{\mu})}{(4\pi)^2}
    + \mathcal{O}(g^4).
    \label{eqn:CBfact}
\end{align}
For bottom quarks, the physical spin-dependent
potential is obtained by simply scaling $\Phi^{\text{int/self}}(r,\bar\mu=M_b)$ with the factor
$c_B^2(\bar\mu,M_b)/4M_b^2$, since we have already obtained the potential performed at a scale $\bar\mu= M_b$.
For the charm quark, however, an additional RG running from the bottom quark mass down 
to the charm quark mass $M_c$ is required,  using the RG factor,
$\exp\left[-\gamma_0 \int_{\bar\mu=M_b}^{\bar\mu=M_c}2g^2(\bar\mu)/\bar\mu\,d\bar\mu\right]$. 
The physical spin-dependent potential is then obtained by multiplying the factor $c_B^2(\bar\mu=M_c,M_c)/4 M_c^2$, 
evaluated at the charm mass. 
For the charm quark, we use the pole mass $M_c=1.35~\text{GeV}$; details of this choice are given in \cref{app1}. This procedure amounts to renormalizing the potential by multiplicative factors $Z_b \simeq 1.23$ and $Z_c \simeq 1.32$ 
for bottom and charm quarks, respectively. 

\begin{figure}[h]
    \centering
    \includegraphics[width=0.45\textwidth]{ 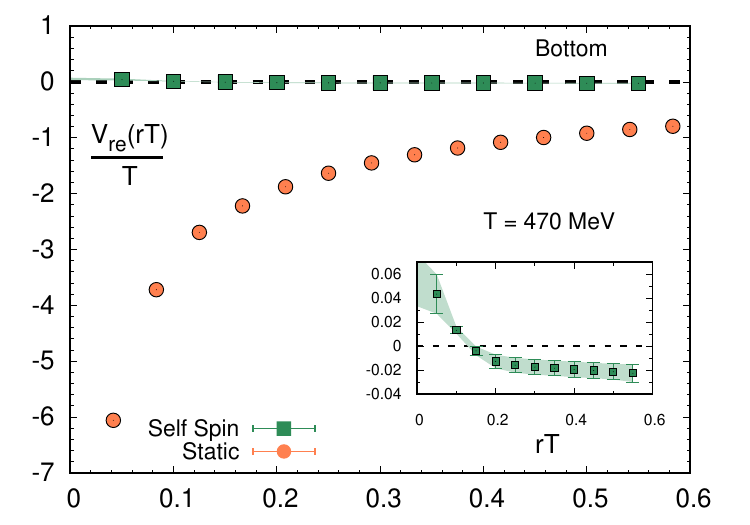}
    \includegraphics[width=0.45\textwidth]{ 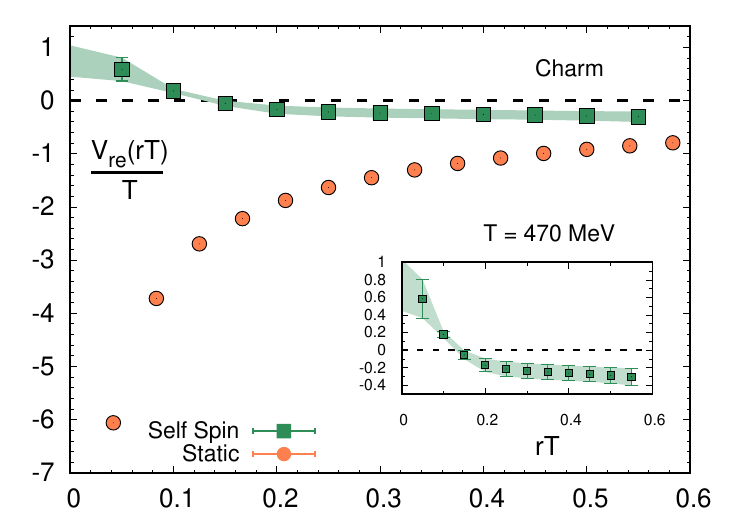}
    \caption{The real part of the self part of the physical spin-dependent potential as a function of $rT$ is compared with the corresponding real part of the static thermal potential at $T=470$ MeV, for bottom (left) and charm (right) quarks respectively.}
    \label{fig:self_static}
\end{figure}

\begin{figure}[tbp]
    \centering
    \includegraphics[width=0.45\textwidth]{ 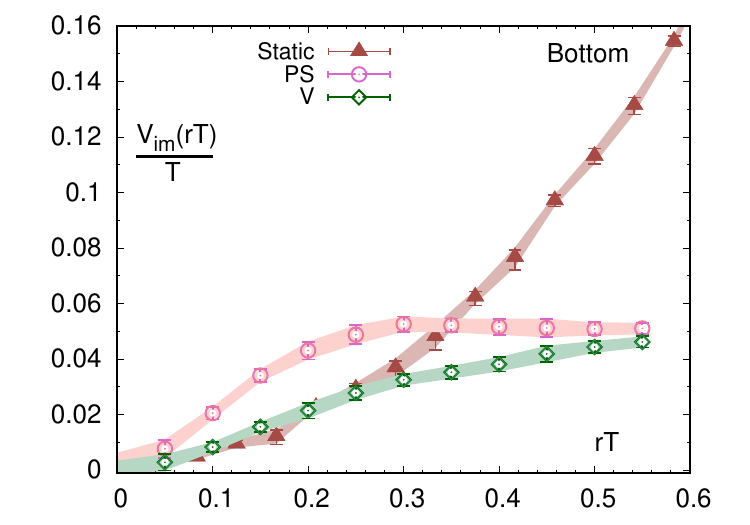}
    \includegraphics[width=0.45\textwidth]{ 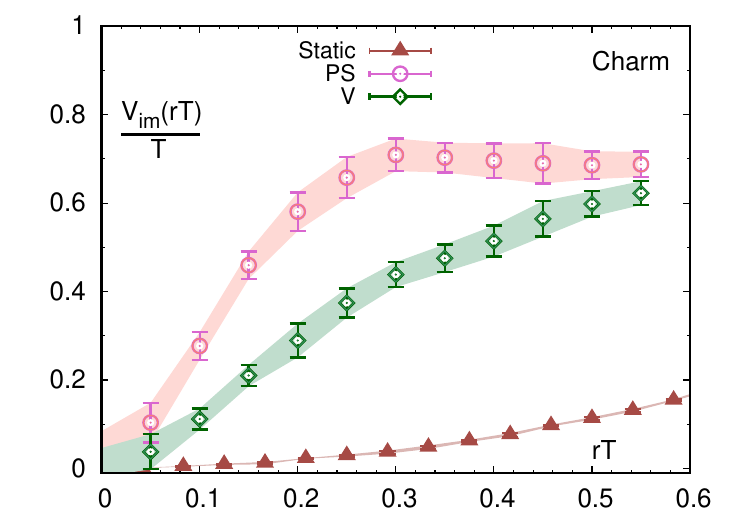}
    \caption{Imaginary part of the static and the spin-dependent potential for the pseudoscalar and vector bottomonium (left panel) and charmonium (right panel) states respectively, shown as a function of $rT$.}
    \label{fig:spin_static_imag}
\end{figure}

The real part of the self contribution of the spin-dependent potential for bottom and charm quarks as a 
function of $rT$ is shown in \cref{fig:self_static}. These values are also 
compared with the real part of the thermal static potential. The details of the calculation of the thermal static potential and the adjustment of the additive constant are discussed in \cref{app1}. One observes that corrections to the 
real part of the static $q\bar q$ potential due to spin interaction are negative 
and subdominant. At large distances, $rT>0.2$, its magnitude saturates to 
$\sim -10~\mathrm{MeV}$ and $\sim -100~\mathrm{MeV}$ for bottom and charm
quarks, respectively. This behavior is also expected from our perturbative
calculations in \cref{eqn:Imagpartpertspinspin}. As already mentioned, 
the short-distance part depends on the resolution scale and therefore is not
quantitatively reliable. However, calculations within pNRQCD suggest that thermal 
corrections should vanish linearly as a function of $r$ at short distance scales. 
This observation will be useful when calculating the quarkonium spectral functions. These results suggest that, while the impact of this constant spin-dependent part on in-medium bottomonium dynamics is strongly suppressed compared to its static part, it can still play a role in charmonium dynamics.

The real part of the interaction part of the spin-dependent potential is consistent with a smeared delta 
function for $rT\gtrsim0.1$, as we have seen in \cref{fig:spin_interacting}. This allows us to take the limit $\epsilon\to 0$ and identify the usual contact term in the real part of the spin-dependent potential, given by
\begin{equation}
\frac{V^{\text{spin}}_{\mathrm{re}}}{T}=
Z_q \frac{32\pi\alpha(\mu=M_b)}{9M^2T}\,\vec{s}_1\cdot\vec{s}_2\,\delta^{(3)}(\vec r).
\label{eqn:spin_int}
\end{equation}
It is important to remind here that there can be additional corrections to this contact potential due to thermal effects. However, these corrections are proportional to $r$ at short distances, as evident from pNRQCD, and are subdominant, thus difficult to extract from our lattice data.

The imaginary part of the spin-dependent potential is obtained by combining the two 
different contributions according to \cref{defpot}. The results for pseudoscalar and vector channel are shown in \cref{fig:spin_static_imag}, separately for bottomonium (left panel) 
and charmonium (right panel) states. As a comparison, we also show the imaginary part of the static potential 
in the same figure as discussed in \cref{app1}. First, the correction to the static potential 
due to spin interactions is significantly larger for charmonium compared to bottomonium states, saturating 
at large distances $rT>0.5$. For bottom quarks, the correction term becomes subdominant compared to 
the static potential for $rT \gtrsim 0.3$, whereas for charm it is expected to happen at $rT \gtrsim 1.2$ 
from a naive extrapolation of the available data. At short distances, the imaginary component of the 
spin-dependent potential in the pseudoscalar channel shows a rapid rise compared to the vector channel, 
eventually approaching each other at distances $rT>0.5$. This short-distance behavior at $rT\ll 1$ can 
be understood qualitatively from pNRQCD, discussed in \cref{eqn:pNRQCD_imag}. Within this effective theory, 
the imaginary potential in the pseudoscalar channel is found to be three times larger than that in the vector 
channel. This is due to the fact that spin-dependent interactions lead to the pseudoscalar color-singlet 
state transition to three vector color-octet states, whereas a vector color-singlet state can transition to 
one pseudoscalar color-octet state. At large distances, on the other hand, one expects saturation as the interaction part of 
spin potential approaches zero, which is also qualitatively consistent with \cref{eqn:Imagpartpertspinspin}.
To summarize, our first results for the imaginary part of the spin-dependent potential show that it has a sizeable 
contribution compared to the static imaginary part at distance scales nearly comparable to the size of the quarkonium 
states. This effect is more prominent in charmonium compared to the bottomium states. As a consequence, we expect its 
impact on the spectral properties of quarkonium, which we investigate in the next section by calculating the spectral 
functions.

\subsection{Spectral functions}

\begin{figure}[tbp]
    \centering    
   \includegraphics[width=0.45\textwidth]{ 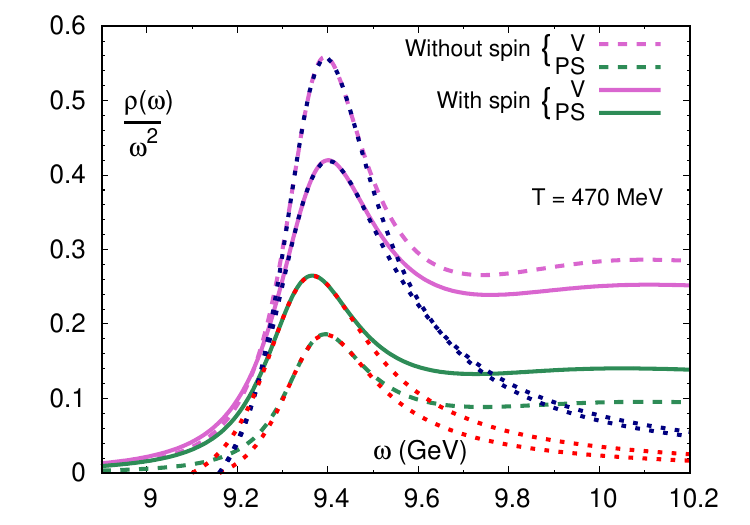}
    \includegraphics[width=0.45\textwidth]{ 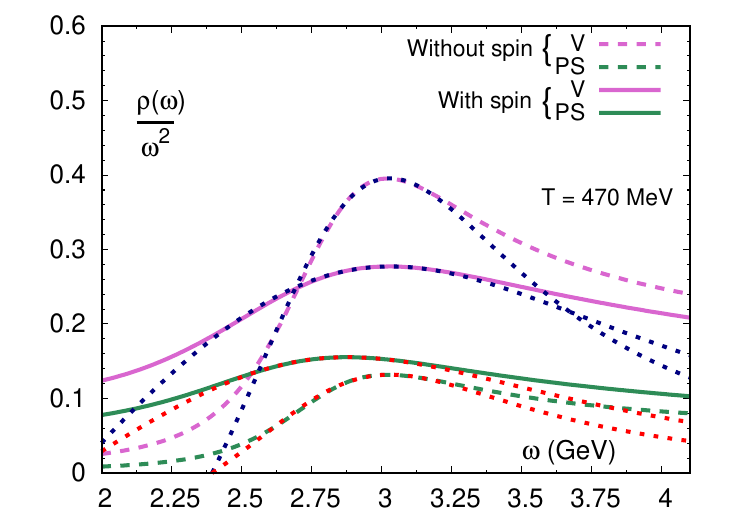}
    \caption{The spectral functions for bottomonium (left) and charmonium (right) states, shown in the absence (dashed lines) and presence (solid lines) of the spin-dependent potential. The dotted line represents the fit using the form given in \cref{eqn:BW}.}
    \label{fig:spectral_functions}
\end{figure}

We will now study the impact of the spin-dependent potential on quarkonium states by calculating their spectral 
functions in the vicinity of the bound states. In this region, the time dependence of the thermal expectation 
value of the following correlation function, representing a static quark-antiquark pair, 
\begin{equation}
C^{>}_{\Gamma}(\vec r,\vec r^{\prime},t)=\int d^3 x\, \left\langle \theta^{\dagger}(\vec x+\vec r/2,t) \Gamma \phi(\vec x-\vec r/2,t)    \theta(\vec x+\vec r^{\prime}/2,0) \Gamma \phi^{\dagger}(\vec x-\vec r^{\prime}/2,0) \right\rangle_{T}.
\end{equation}
can be described by a Schrödinger equation in the presence of the thermal potential,
\begin{equation}
\left[2M-\frac{\nabla_r^2}{M}+V_{\Gamma}(r,T)\right]C^{>}_{\Gamma}(\vec r,\vec r^{\prime},t)=i\frac{\partial C^{>}_{\Gamma}(\vec r,\vec r^{\prime},t)}{\partial t}~.
\end{equation}
The above equation is solved with an initial condition 
$C^>(\vec r,\vec r^{\prime},0)=\lambda \delta^3(\vec r-\vec r^{\prime})$, where $\lambda=6\,N_c$ 
for vector and $\lambda=2\,N_c$ for pseudoscalar channel, respectively \cite{Burnier:2007qm}.
Note that the source term at $t=0$ already contains explicit spin dependence. This approach has been used for 
spectral function reconstruction in earlier studies, employing both perturbative and non-perturbative static 
potentials \cite{Burnier:2007qm, Ali:2025iux} but without the spin-dependent potential. The details of the calculation of the 
spectral function in the presence of spin interactions are given in \cref{app1}. The resulting spectral 
functions are shown in \cref{fig:spectral_functions}. The dashed (solid) lines represent the spectral functions without
(with) the inclusion of the spin-dependent potential. In the absence of the spin-dependent potential, the difference between the pseudoscalar and vector spectral functions  arises purely from the trivial spin dependence of the source term.  Even in the absence of 
spin interactions at $T=470$ MeV, only the $1S$ bottomonium survives with an observable thermal broadening. 
In contrast, the $1S$ charmonium spectral function has a much larger thermal broadening, raising concerns about whether it 
could even be associated with a well-defined physical state.

The spin interaction breaks the trivial degeneracy arising due to the choice of the source term by introducing 
additional dynamical effects. From \cref{fig:spectral_functions} it is evident that the spectral weight in the vector 
channel is suppressed near the bound-state region, whereas in the pseudoscalar channel it is enhanced. The vector 
spectral function is related to the dilepton production rate in 
QGP \cite{PhysRevD.31.545}. Suppression of the vector spectral function due to the spin interactions can thus play an important role in the 
physics of quarkonium polarization observed in heavy-ion collision experiments. 

To quantify the effect of spin interactions on the decay width of quarkonium spectral functions, we fit them 
near the peak at $\omega_p$, using the following skewed Breit-Wigner ansatz \cite{Burnier:2015tda},
\begin{equation}
\rho(\omega \approx \omega_p)
=
A\,\frac{\Gamma/2}{(\omega-\omega_p)^2+\Gamma^2/4}
+
\delta\,\frac{(\omega-\omega_p)\Gamma}{(\omega-\omega_p)^2+\Gamma^2/4}
+\cdots .
\label{eqn:BW}
\end{equation}
The results of the fit are shown in \cref{fig:spectral_functions} by the dotted lines. 
For bottomonium, the decay width $\Gamma \simeq 250~\mathrm{MeV}$ obtained in the absence of spin interactions becomes $\Gamma \simeq 280~\mathrm{MeV}$ in the pseudoscalar channel and $\Gamma \simeq 310~\mathrm{MeV}$ in the vector channel due to spin interactions.
 Thus, the pseudoscalar state has a $\sim 10\%$ 
smaller width than the vector state, even though the imaginary part of the spin-dependent potential 
is larger in the pseudoscalar channel. This can be understood from the fact that the real part of the spin 
interaction potential makes the pseudoscalar state more strongly bound than the vector state, compensating the effects due to the larger imaginary component. The total decay width is thus due to 
the competition between these two effects. 
The width of charmonium states is already large, $\Gamma \simeq 1275~\mathrm{MeV}$ in the absence of spin 
interactions. Such a large width indicates that the charmonium $1S$ state is not a well-defined bound state 
at this temperature. Including spin interactions further increases the width of the state, resulting in $\Gamma \simeq 2261~\mathrm{MeV}$ in the pseudoscalar channel and $\Gamma \simeq 2669~\mathrm{MeV}$ in the vector channel.  The increase of 
the widths further ensures that the charmonium $1S$ state cannot survive in the QGP medium at this temperature.

\section{Summary \& Outlook}

In this work, we have calculated the spin-dependent potential between a static quark-antiquark pair in quenched 
QCD at a temperature, $T=470~\mathrm{MeV}$, from Wilson-line correlators with chromomagnetic insertions using lattice 
techniques. Our most important finding is that the thermal spin-dependent potential is a complex quantity, 
similar to the thermal static potential. The calculation of the spin-dependent potential thus also requires an analytic 
continuation from Euclidean to real time. This is an ill-posed problem unless additional physical inputs are provided. 
Guided by the perturbative expression of the correlator and requiring the existence of a spin-dependent potential of the form defined in \cref{defpot}, we used the ansatz \cref{eqn:Param_spin} to extract the potential. We extracted the potential from a continuum-estimated renormalized correlator.

The real part of the self spin-dependent potential, arising due to color-magnetic field insertions on the same Wilson line, results in a thermal shift of $\sim -10$ MeV for bottomonium states. This shift is significantly larger by a factor of $10$ for charmonium states, as shown in \cref{fig:self_static}. The real part of the interaction part of the spin-dependent potential between a heavy quark-antiquark pair is calculated from the Wilson line correlator, but now with color-magnetic field insertions on different Wilson lines. We found that this interacting spin-dependent potential can be described by the regularized Dirac delta function given in \cref{eqn:ftdelta}.
 Its imaginary part exhibits even more interesting features. Its magnitude is zero at $r=0$, reaches a maximum around $rT\sim 0.2$ and approaches zero again at large quark-antiquark separations. This is unlike the imaginary component of the self spin-dependent potential, which starts from a vanishing value at $r=0$ and saturates at large distance. Combining these results for the potential in the pseudoscalar and vector channels for charm quarks, we observe that at distances $rT \lesssim 1.2$, its magnitude for charmonium states dominates over its counterpart in the static case. Similarly, for bottom quarks at distances $rT \lesssim 0.38$, the magnitude for bottomonium states dominates over its counterpart in the static case. The difference between them vanishes when $r\to 0$, as depicted in \cref{fig:spin_static_imag}.

We have also discussed the physical implications of these findings by calculating the spectral functions in the 
pseudoscalar and vector channels for both charmonium as well as bottomonium.  We observe that spin interactions enhance the thermal decay widths of the quarkonium states. The magnitude of this enhancement in bottomonium is about $\sim 12~\%$ for the pseudoscalar state and $\sim 24~\%$ for the vector state. In the case of charmonium, the enhancement increases dramatically. The 
spin-dependent potential thus aggravates the thermal distortion of charmonium states, ensuring that these do not 
survive within the gluon plasma at these temperatures. In the future, we will report on the impact of spin interactions 
in QCD with physical quark masses, which is a work in progress. This would allow us to explain several 
interesting physical phenomena, e.g., the quarkonium polarization, observed in the heavy-ion collision experiments.

\section{Acknowledgments}
This work was supported by the Deutsche Forschungsgemeinschaft (DFG, German Research Foundation) – project number 
315477589 – TRR 211. Computations have been performed on the GPU Cluster at Bielefeld University. We acknowledge the 
allocation of computational resources for this project. Our codes was in part based on the \texttt{SIMULATeQCD} 
library~\cite{HotQCD:2023ghu}. We would like to thank Saumen Datta for helpful discussions, Rajiv V. Gavai for 
suggestions related to spectral functions and Guy Moore for discussions related to the renormalization 
of correlators. 
This research was supported for D.B., S.S. and S.T., in part, by the International Centre for Theoretical Sciences (ICTS) 
for participating in the program $-$ Hard probes in non-equilibrium QCD matter 2026 (ICTS/NQCD2026/03).

\appendix
\section{Appendix}
\label{app1}
\subsection{Static potential}

We describe the details of the calculation of the thermal static heavy quark-antiquark potential and discuss how we calculate the additive constant in the real part of this potential for performing the comparison shown in \cref{fig:self_static}. To ascertain the additive constant, we also need to calculate the potential at zero temperature. The ensembles used for the calculation of the static potential in quenched QCD correspond to temperatures $T=0.75~T_d$ and $T=1.5~T_d$, representing the confined and deconfined phases of SU(3) respectively, details of which are mentioned in  \cref{tab:paramReV}. 

\begin{table}[h]
\begin{tabular}{|c|c|c|c|c|}
\hline
$T/T_d$ & $N_\tau$ & $\beta$ & $a$ (fm) & $N_\text{configs}$ \\
\hline
$0.75$ & $48$ & $7.192$ & $0.0175$ & $1460$ \\
\hline
$1.5$  & $24$ & $7.192$ & $0.0175$ & $4780$ \\
\hline
\end{tabular}
\captionof{table}{Parameters for the lattice calculation of thermal static potential in quenched QCD.}
\label{tab:paramReV}
\end{table}

\begin{figure}[tbp]
    \centering
    \includegraphics[width=0.45\textwidth]{ 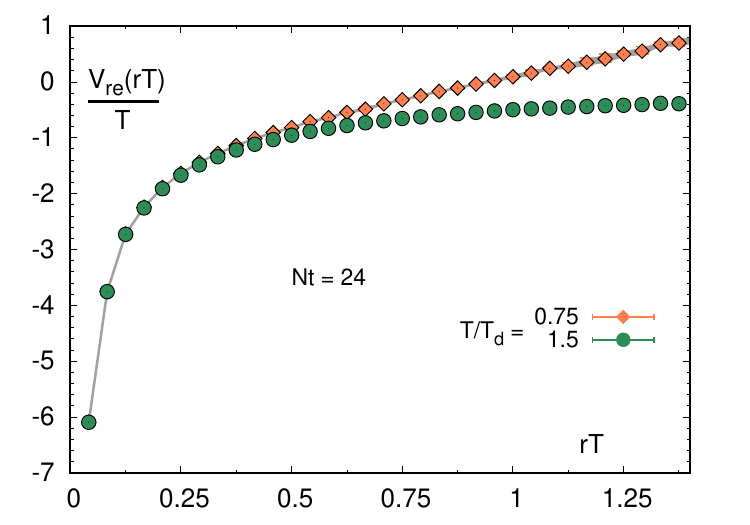}
    \caption{The static quark-antiquark potential at zero ($T\simeq 0.75~T_d$) and finite temperatures ($T\simeq 1.5~T_d$) as a function of $rT$ in quenched QCD, where $T_d$ is the deconfinement transition temperature.}
    \label{fig:static_pot}
\end{figure}

The static potential has been obtained from a pure exponential fit to the Wilson line correlator at $\sim 0.75 ~T_d$, which mimics the zero-temperature scenario as well. The following fit ansatz was used to describe the resulting potential:
\begin{equation}
V(r)=-\frac{\alpha}{r_I}+\sigma r+c
\end{equation}
where $r_I$ denotes the improved definition of lattice distance defined as,
\begin{equation}
\frac{1}{r_I(\vec n)}=4\pi \int_{-\pi}^{\pi} \frac{d^3 q}{(2\pi)^3} \frac{\cos(\vec q\cdot \vec n)}
{4\sum_{i=1}^{3}\sin^2(q_i/2)}.
\end{equation}
In this way of defining the distance, the short-distance lattice artifacts are reduced \cite{Sommer:1993ce}, where $\vec r=a\vec n$ 
denotes the separation between the static heavy $q\bar q$ pair. The additive constant $c$ is divergent in the 
continuum limit. We estimate this constant by requiring that the spin-averaged $1S$ bottomonium mass, 
$\frac{1}{4}M_{\eta_b(1S)}+\frac{3}{4}M_{\Upsilon(1S)}=9.4449~\mathrm{GeV}$, agrees with the experimental value 
reported in Ref.~\cite{ParticleDataGroup:2024cfk}. We use the bottom quark mass $M_b=4.78~\mathrm{GeV}$. Having 
obtained this additive constant, we tune the charm quark mass such that the experimentally measured value of the 
spin-averaged $1S$ charmonium mass, $\frac{1}{4}M_{\eta_c(1S)}+\frac{3}{4}M_{J/\psi(1S)}=3.0687~\mathrm{GeV}$, 
is reproduced. This gives us a charm mass $M_c=1.35~\mathrm{GeV}$. The thermal potential has been obtained by 
performing a fit to the Wilson line correlator using \cref{eqn:Param_WL}. Subtracting the same additive term 
from the bare finite-temperature potential, the real part of the resulting renormalized potential is shown in 
\cref{fig:static_pot}. We have also compared between the imaginary part of this static potential with the 
imaginary component of the spin-dependent potential in \cref{fig:spin_static_imag}.

\subsection{Computing the spectral function}
In order to calculate the spectral function $\rho(\omega)$ for a quarkonium 
state we have to first determine the correlation function $\Psi(\omega;\vec r,\vec r')$
from the following equation~\cite{Burnier:2007qm},
\begin{equation}
\left[
\omega-2 M +\frac{\nabla^2}{M} - V(r)-\delta V(r)
\right]
\Psi(\omega;\vec r,\vec r')
=
\lambda \,\delta^{(3)}(\vec r-\vec r') .
\label{eq:QuarkoniumCorrEvol}
\end{equation}
where $\lambda$ depends on the spin structure of the state. The term $\delta V(r)=c\,\delta^3(\vec r)$ denotes the 
contact term arising from the spin interaction. All the remaining 
contributions to the potential, namely the real and imaginary parts of the static 
potential, the imaginary part of the spin-dependent potential, and the self 
contribution to its real part, are included in $V(r)$.
The spectral function is then obtained in the limiting case such that 
$\rho(\omega)=-\lim_{r\to0,r^{\prime}\to 0}\operatorname{Im}\Psi(\omega;\vec r,\vec r')$.
If  $G_0(\omega,\vec r,\vec r')$ is the Green's function in the absence of $\delta V$
then  $\Psi(\omega,\vec r,\vec r^{\prime})$ can be solved iteratively using~\cref{eq:QuarkoniumCorrEvol},
\begin{equation}
\begin{aligned}
\Psi(\omega,\vec r,\vec r')
&=
\lambda G_0(\omega,\vec r,\vec r')
+
\lambda \int d^3 \vec r_1\,
G_0(\omega,\vec r,\vec r_1)\,
\delta V(r_1)\,
G_0(\omega,\vec r_1,\vec r')
\\
&\quad
+
\lambda \int d^3 \vec r_1\, d^3 \vec r_2\,
G_0(\omega,\vec r,\vec r_1)\,
\delta V(r_1)\,
G_0(\omega,\vec r_1,\vec r_2)\,
\delta V(r_2)\,
G_0(\omega,\vec r_2,\vec r')
+\cdots .
\end{aligned}
\end{equation}
Substituting $\delta V(r)$ one obtains the following resummed expression of the correlation function,
\begin{equation}
\Psi(\omega,\vec r,\vec r\,')
=
\lambda\left(G_0(\omega,\vec r,\vec r\,')
+
\frac{
c\,G_0(\omega,\vec r,\vec 0)\,
G_0(\omega,\vec 0,\vec r\,')
}{
1-c\,G_0(\omega,\vec 0,\vec 0)
}\right)~.
\end{equation}
The spectral function $\rho(\omega)$ is then simply,
\begin{equation}
\rho(\omega)
=
-\operatorname{Im}\Psi(\omega,\vec 0,\vec 0)
=
-\lambda \operatorname{Im}
\left[
\frac{G_0(\omega,\vec 0,\vec 0)}
{1-c\,G_0(\omega,\vec 0,\vec 0)}
\right]
=
-\frac{\lambda G^{\rm im}_{0}(\omega,\vec 0,\vec 0)}
{\Big[1-c\,\,G^{\rm re}_{0}(\omega,\vec 0,\vec 0)\Big]^2+\Big[c\,G^{\rm im}_{0}(\omega,\vec 0,\vec 0)\Big]^2}.
\end{equation}

The $G^{\rm re}(\omega=0,\vec 0,\vec 0)$ that appears in the denominator has an ultraviolet divergence. As a result, the spectral function $\rho(\omega)$ needs to be renormalized, which can be achieved through the running of the couplings in the potential $\delta V$. The most general interaction can be taken to be $\delta V(r)=(c_0+c_s\,\vec s_1\cdot\vec s_2)\delta^3(\vec r)$, with the couplings running according to
\begin{equation}
\frac{1}{c_0(\Lambda)+c_s(\Lambda)\,\vec s_1\cdot\vec s_2}
=\frac{1}{c_{\rm phys}\,\vec s_1\cdot\vec s_2}+D(\Lambda).
\end{equation}

Here $c_{\rm phys}$ determines the spin splitting between physical singlet and triplet states, while 
$D(\Lambda)$ represents the divergent contribution at a cutoff $\Lambda$. Using this relation, we obtain 
the renormalized spectral function in terms of physical couplings as,
\begin{equation}
\rho_r(\omega)
=
-\frac{\lambda\, G^{\rm im}_{0}(\omega,\vec 0,\vec 0)}
{\Big[1-c_{\text{phys}}\,\vec s_1\cdot\vec s_2\,G^{\rm re}_{0}(\omega,\vec 0,\vec 0)_{\text{sub}}\Big]^2+\Big[c_{\text{phys}}\,\vec s_1\cdot \vec s_2\,G^{\rm im}_{0}(\omega,\vec 0,\vec 0)\Big]^2}.
\label{eqn:fitfunction}
\end{equation}
We have defined $ G^{\rm re}_0(\omega,\vec 0,\vec 0)_\text{sub} = G^{\rm re}_0(\omega,\vec 0,\vec 0)-D$ and $D$ is 
chosen such that it cancels the divergent term in the bare Green's function, $G^{\rm re}_0(\omega,\vec 0,\vec 0)$. 
We have found that $G^{\rm re}_0(\omega,\vec 0,\vec 0)$ is a constant as a function of $\omega$, except near the 
threshold and bound-state regions, where nontrivial physical structures appear.  This large constant piece is precisely 
the contribution that we would like to remove from the spectral function in order to obtain the physical states; 
as a result, we choose $D=G^{\rm re}_0(\omega=0,\vec 0, \vec 0)$. The renormalized spectral function obtained as a result of this subtraction is shown in \cref{fig:spectral_functions}, where the subscript $r$ has been omitted.

For the computation of the spectral function, we have taken $c_{\text{phys}}$ from \cref{eqn:spin_int}.
We need a parametric form of the potential for all the potentials in order to solve the Schrödinger equation in \cref{eq:QuarkoniumCorrEvol}. 
At zero temperature, we use the Cornell potential, with its coefficients 
determined from the improved-distance fit described above. At finite temperature, we 
use an interpolated potential. For the real part, at short distances, $rT \lesssim 0.3$, where we find the potential to be 
temperature independent, we replace it with the zero-temperature Cornell 
potential. This replacement reduces the short-distance cutoff artifacts in the 
real part of the potential, since the latter incorporates the improved-distance correction.

The short-distance part of the real 
component of the self-spin interaction potential is obtained by performing a linear extrapolation of the 
potential data to zero from $rT \lesssim 0.2$. This choice is motivated by pNRQCD effective theory, where 
the finite temperature contribution vanishes linearly at short distances. 
As the real part of the self potential, the 
imaginary part of the pseudoscalar and vector channels becomes nearly constant at 
long distances, we have taken their values to be constant beyond the available data points. For the imaginary part of the 
static potential, we have taken a constant value at very large distances, which is 
motivated by perturbation theory calculations.

\bibliography{PaperPRDSW.bib}

\end{document}